\documentclass[aps,prl,twocolumn,showpacs,superscriptaddress]{revtex4-1}
\usepackage{latexsym}
\usepackage{amssymb}
\usepackage{graphicx}
\usepackage{amsmath}
\usepackage{bm}
\usepackage[colorlinks,
          linkcolor=black,
            citecolor=black,
            urlcolor=blue
           ]{hyperref}
\usepackage{verbatim}
\usepackage{mathrsfs}
\usepackage{extarrows}
\usepackage{comment}
\usepackage{mathtools,slashed}
\usepackage{soul}
\usepackage[toc,page]{appendix}
\usepackage[vcentermath]{youngtab}
\usepackage{multirow}
\usepackage[width=.75\textwidth]{caption}
\usepackage{atbegshi,picture}
\usepackage{lipsum}

\makeatletter
\renewcommand\@makecaption[2]{
  \par
  \vskip\abovecaptionskip
  \begingroup
   \small\rmfamily
    \begingroup
     \samepage
     \flushing
     \let\footnote\@footnotemark@gobble
     \@make@capt@title{#1}{#2}\par
    \endgroup
  \endgroup
  \vskip\belowcaptionskip
}
\makeatother

\AtBeginShipoutNext{\AtBeginShipoutUpperLeft{
  \put(\dimexpr\paperwidth-1cm\relax,-1.5cm){\makebox[0pt][r]{IPMU-18-0086}}
}}

\makeatletter\begin{document}

\title{Anomaly matching and symmetry-protected critical phases in $SU(N)$ spin systems in 1+1 dimensions}

\author{Yuan Yao}         
\thanks{The first two authors contributed equally to the work.}
\affiliation{Institute for Solid State Physics, The University of Tokyo. Kashiwa, Chiba 277-8581, Japan}

\author{Chang-Tse Hsieh}
\thanks{The first two authors contributed equally to the work.}
\affiliation{Kavli Institute for the Physics and Mathematics of the Universe (WPI),
The University of Tokyo Institutes for Advanced Study,
The University of Tokyo, Kashiwa, Chiba 277-8583, Japan}
\affiliation{Institute for Solid State Physics, The University of Tokyo. Kashiwa, Chiba 277-8581, Japan}

\author{Masaki Oshikawa}
\affiliation{Institute for Solid State Physics, The University of Tokyo. Kashiwa, Chiba 277-8581, Japan}

\date{\today}

\begin{abstract}

We study $(1+1)$-dimensional $SU(N)$ spin systems in the presence of the global $SU(N)$ rotation and lattice translation symmetries. By matching the mixed anomaly of the $PSU(N)\times\mathbb{Z}$ symmetry in the continuum limit, we identify a topological index for spin model evaluated as the total number of Young-tableaux boxes of spins per unit cell modulo $N$, which characterizes the ``ingappability'' of the system.
A nontrivial index implies either a ground-state degeneracy in a gapped phase, which can be regarded as a field-theory version of the Lieb-Schultz-Mattis theorem, or a restriction of the possible universality classes in a critical phase -- the symmetry-protected critical phase, e.g. only a class of $SU(N)$ Wess-Zumino-Witten theories can be realized in the low-energy limit of the given lattice model in the presence of the symmetries.
Similar constraints also apply when a higher global symmetry emerges in the model with a lower symmetry. 
Our prediction agrees with several examples known in previous studies of $SU(N)$ models.

\end{abstract}

\maketitle

\textit{Introduction.---}
{
Identification of the phase of a many-body quantum system is an important but in general hard problem in condensed matter physics.
Quite often, symmetries play an essential role in determining the phase.
As a notable example, Lieb-Schultz-Mattis (LSM) theorem and its generalizations~\cite{Lieb:1961aa, Affleck:1986aa,OYA1997,Oshikawa:2000aa, Hastings:2004ab} state that a lattice model cannot be a trivial symmetric insulator if the filling per unit cell is fractional and if the translation symmetry and particle number conservation are strictly imposed.
This gives a strong constraint on possible gapped phases realized
in a given microscopic model.
On the other hand, classification of critical phases is also an important problem.
In addition to the quantum critical points between gapped phases,
many stable critical phases have been found numerically and experimentally~\cite{Alcaraz:1989aa,Fuhringer:2008aa,Matsumoto-YbAlB4_Science2011}.
However, the reason of such stability is not completely understood. 
Moreover, much less is known about universal constraints on the critical phases,
while a related proposal has been addressed in $SU(2)$ spin chains recently~\cite{Furuya:2017aa}.

Systems with global symmetries higher than the conventional $U(1)$ or $SU(2)$,
in particular the $SU(N)$ symmetry are of intense interest.
For a long time, $SU(N)$ symmetric systems have been studied as a theoretical toy model to understand the ``physical'' $SU(2)$ spin systems.
Recently, moreover, ``spin'' systems with an $SU(N)$ symmetry with $N>2$ 
are realized in ultracold atoms
on optical lattices~\cite{Wu:2003aa,Honerkamp:2004aa,Cazalilla:2009aa,Gorshkov:2010aa,Taie:2012aa,Pagano:2014aa,Scazza:2014aa,Zhang:2014aa}.
A realization in spin-orbital systems is also suggested~\cite{YamadaSU4}. 
Thus the study related to the phase diagrams of $SU(N)$ spin systems is of realistic interest in its own. 
Furthermore, 
even when only the $SU(2)$ spin-rotation and translation symmetries are imposed, they can be enhanced to higher symmetries.
For example, the spin-$1$ bilinear-biquadratic chain has an explicit $SU(3)$ symmetry at a special point called as the Uimin-Lai-Sutherland model~\cite{Uimin:1970aa,Lai:1974aa,Sutherland:1975aa}. 
Higher symmetries can also emerge in the thermodynamic limit, even if the microscopic model does not have such symmetries exactly.
For example, an emergent $SU(3)$ symmetry is found in a critical spin-$2$ chain~\cite{Chen:2015aa}. 
It would be important to find universal constraints on such symmetry enlargements.

In this Letter, we focus on fundamental constraints on the phase diagrams of $(1+1)$d $SU(N)$ spin systems with spin-rotation symmetry and lattice translation symmetry.
Our approach is based on the idea of 't Hooft anomaly matching~\cite{tHooft:1980xss}, which has been used to investigate infrared behaviors of strongly coupled systems (for examples, see recent works~\cite{Benini:2017aa, Gaiotto:2017aa, Tanizaki:2017aa, Komargodski:2017aa, Komargodski:2018aa, Shimizu:2017aa}). 
Specifically, we identify a topological quantity for a lattice spin model that matches the ``anomaly'' of the relevant symmetries in the low-energy phases and is given by the total number of Young-tableau boxes of spins per unit cell modulo $N$.
It predicts whether the system admits a unique gapped ground state and further restricts ground-state degeneracies (GSDs) in gapped phases. 
Our result agrees with the $SU(N)$ generalization of the LSM theorem~\cite{Affleck:1986aa}.
In addition, it imposes constraints on possible critical phases,
going beyond the scope of the LSM-type theorems.
That is, we postulate a classification of symmetry-protected critical (SPC) phases regarding the Wess-Zumino-Witten (WZW) universality classes of the $SU(N)$ spin models based on the anomaly matching, generalizing the proposal for the $SU(2)$ case~\cite{Furuya:2017aa}.

Moreover, we also obtain a constraint on a higher $SU(N')$ symmetry with $N'>N$ in a model with the $SU(N)$ symmetry only by matching their symmetry anomaly.
As a special example, such restriction can explain that $SU(3)$ symmetry 
enlargement
has not (actually cannot) been found in $SU(2)$ and translation-symmetric half-integer spin chains. 
This demonstrates the power of our anomaly-based approach, and would pave a new way to discuss symmetry enlargements in general.

\textit{Translationally invariant $SU(N)$ spin system in $1+1$ dimensions and the LSM index ---}
We consider a generic $(1+1)d$ $SU(N)$ spin system described by a Hamiltonian $\mathscr{H}_{SU(N)}$ with the translation symmetry and an $SU(N)$ spin-rotation symmetry specified by the projective special unitary group $PSU(N)$. 
Here the system is subject to periodic boundary condition and the translation defines a unit cell consisting of, for  generality, multiple spins
and forms a discrete group $\mathbb{Z}^{\mathrm{trans}}$ in the thermodynamic limit. 
A typical example of such a system is the $SU(N)$ Heisenberg antiferromagnetic (HAF)
model
\begin{eqnarray}
\label{heisenberg}
\mathscr{H}_{\mathrm{HAF}}=J\sum_{\langle i,j\rangle,\alpha,\beta}S^\alpha_{i,\beta}S^\beta_{j,\alpha}, 
\quad J > 0.
\end{eqnarray}
where $\alpha$ and $\beta$ are the spin indices that take values among $1$ to $N$ and the $SU(N)$ generators satisfy the following $su(N)$ Lie algebra commutation relations: 
\begin{eqnarray}
\left[S^\alpha_{i,\beta},S^\gamma_{j,\delta}\right]=\delta_{i,j}\left(\delta^\alpha_\delta S^\gamma_{i,\beta}-\delta^\gamma_\beta S^\alpha_{i,\delta}\right). 
\end{eqnarray}
The first question to be asked is whether $\mathscr{H}_{SU(N)}$ has a trivial symmetric gapped ground state (a translationally invariant $SU(N)$ singlet). 
Here we give an approach to answering this question, based on the idea of symmetry anomalies.

The microscopic symmetry group $PSU(N)\times\mathbb{Z}^{\mathrm{trans}}$ of the spin system, which incorporates both on-site and non-on-site symmetries, transmutes to a purely internal symmetry group $PSU(N)\times\mathbb{Z}$ in the continuum effective field theory (EFT) that describes the system in the low-energy limit.
While both $\mathbb{Z}^{\mathrm{trans}}$ and $\mathbb{Z}$ are the infinite cyclic group, we emphasize the non-on-site nature of the lattice translations by denoting $\mathbb{Z}^{\mathrm{trans}}$.
In the EFT, there is possibly a conflict between the internal $PSU(N)$ and $\mathbb{Z}$ symmetries: they might have a mixed 't Hooft anomaly, that is, gauging one symmetry causes the breakdown of the other 
\footnote{
For example, when the system is in a gapless phase, there are both left- and right-moving soft (chiral) modes present at low energy, and the $PSU(N)$ and $\mathbb{Z}$ symmetries would act on these modes as a ``vector'' and an ``axial'' symmetries, respectively. In this situation, there is a potential chiral/axial anomaly in the low-energy field theory, similar to the case of the $(1+1)$d Dirac theory with both a vector and an axial $U(1)$ symmetries.}.
An example of such a mixed $PSU(N)-\mathbb{Z}$ anomaly is the phase ambiguity of the partition function of a low-energy EFT, e.g. the $SU(N)$ WZW  conformal field theory (CFT) (which we will discuss later), under a $\mathbb{Z}$ transformation after coupled to a background $PSU(N)$ gauge field.
This type of anomaly of the EFT can be traced back to the non-on-site nature of the lattice translation in the microscopic lattice model~\cite{Cheng:2016aa, Jian:2018aa, Cho:2017aa, Metlitski-Thorngren17, Cheng:2018aa}, and there should be, based on the concept of 't Hooft anomaly matching, a quantity that can be determined from the underlying lattice model and matches the mixed anomaly in the continuum limit. We call such a quantity as the \textit{LSM index} $\mathcal{I}_{N}$:
\begin{align}
\label{anomaly-matching}
\text{LSM index } \mathcal{I}_{N} \longleftrightarrow \text{mixed $PSU(N)-\mathbb{Z}$ anomaly}.
\end{align}

The LSM index defined above is a topological quantity that is independent of the underlying spin interactions and characterizes the ``ingappability'' of a spin system. That is, any $(1+1)$d spin model with a nontrivial index is ``ingappable'' as long as both $PSU(N)$ and $\mathbb{Z}^{\mathrm{trans}}$ symmetries are respected: the system can be in either a symmetric gapless phase or a phase with spontaneous symmetry breaking, e.g. a $\mathbb{Z}^{\mathrm{trans}}$-broken gapped phase or a $PSU(N)$-broken gapless phase
\footnote{
The $PSU(N)$-broken gapless phase is actually the ferromagnetic phase, which exists only at zero temperature and does not survive at finite temperature, according to the Mermin-Wagner Theorem.
}. 
This is implied by the relation between symmetry anomalies (at low energy) and ingappability, which has been studied on the boundary theories of symmetry-protected topological (SPT) phases
\cite{Ryu:2012ac, Wen:2013aa, Wang:2013aa, Hsieh:2014aa, Hsieh:2016aa, Witten:2016ab}.

\begin{table*}[t]
\centering
\begin{tabular}{c c c c c c }
\hline\hline
Model& \qquad YT \ \ \quad\quad&$\mathcal{I}_{N}$&$\ \quad\mathrm{GSD}\  \quad$&IR CFT; $m$&Mixed anomaly\\
\hline
$SU\!(3)$ trimer model~\cite{Greiter:2007ac}&$\yng(1)$&$1\!\!\mod\!3$&$3\in3\mathbb{N}$&-&-\\
\\[-1em]
$SU\!(3)$ {\bf 10}-VBS model~\cite{Greiter:2007ac}&$\yng(3)$&$0\!\!\mod\!3$&$1\in1\mathbb{N}$&-&-\\
\\[-0.7em]
$SU\!(6)$ {\bf 70}-VBS model~\cite{Greiter:2007aa}&$\yng(2,1)$&$3\!\!\mod\!6$&$2\in2\mathbb{N}$&-&-\\
\\[-0.9em]
$S$-$3/2$ TB\! model\cite{Takhtajan:1982aa,Babujian:1982aa}&$\yng(3)$&$1\!\!\mod\!2$&-&$SU(2)_3$ WZW; $1$&$1\mod2$\\
\\[-0.7em]
$\mathscr{H}^{[3,2]}$ AJ model\cite{Andrei:1984aa,Johannesson:1986aa}&$\yng(2)$&$2\!\!\mod\!3$&-&$SU(3)_2$ WZW; $1$&$2\mod3$\\
\\[-0.7em]
$SU\!(3)$ \!$1\!\times\!2$-\!YT HAF\cite{Rachel:2009aa,Lajko:2017aa}&$\yng(2)$&$2\!\!\mod\!3$&-&$SU(3)_1$ WZW; $2$&$2\mod3$\\
\\[-0.7em]
$SU\!(9)$ \!$2\!\times\!1$-\!YT HAF\cite{Dufour:2015aa}&$\yng(1,1)$&$2\!\!\mod\!9$&-&$SU(9)_1$ WZW; $2$&$2\mod9$\\
\\[-1em]
$SU\!(3)$ $2$-leg ladder~\cite{Lecheminant:2015aa}&$\yng(1)\otimes\yng(1)$&$2\!\!\mod\!3$&-&$SU(3)_1$ WZW; $2$&$2\mod 3$\\
\\[-1em]
\hline\hline
\end{tabular}
\caption{Examples of gapped and critical $SU(N)$ spin systems. For the first three gapped exactly solvable models, the actual GSDs are consistent with our constraint~(\ref{GSD}). For the following critical models, the numerically proposed IR CFTs in the fifth column obey SPC classification specified by~(\ref{matching_LSM_WZW}). VBS: Valence-bond-solid; TB: Takhtajan-Babujian; AJ: Andrei-Johannesson.}
\label{example}
\end{table*}

In order to identify the mixed anomaly -- and hence the LSM index -- in $(1+1)$d systems, it is useful to 
introduce related $(2+1)$d SPT
phases via boundary-bulk correspondence.
An ingappable $(1+1)$d systems in the presence of the symmetries may be regarded as the edge state of a nontrivial $(2+1)$d SPT phase.
The (2+1)d bosonic SPT phases protected by an internal $PSU(N)\times\mathbb{Z}$ symmetry
are classified by the third cohomology $H^3(PSU(N)\times\mathbb{Z}, U(1))$~\cite{Chen:2013aa}. 
According to  the K\"{u}nneth formula, it can be decomposed as 
\begin{align}
&H^3(PSU(N)\times\mathbb{Z}, U(1)) 
\nonumber\\
&\cong H^3(PSU(N), U(1)) \oplus H^2(PSU(N), U(1)).
\end{align}
The first factor $H^3(PSU(N),U(1)) \cong \mathbb{Z}$
represents the $(2+1)$d SPT phases protected solely by the $PSU(N)$ symmetry.
Their edge states cannot be realized in purely $(1+1)$d lattice models
and thus are not of our interest here.
On the other hand, the second factor $H^2(PSU(N),U(1)) \cong \mathbb{Z}_N$
represents the $(2+1)$d SPT phases protected by both the $PSU(N)$ and the $\mathbb{Z}$ symmetries, 
and the anomaly manifested in their edge states is exactly the mixed anomaly in the low-energy states of the $(1+1)$d systems we are looking for.

The mixed anomaly of $PSU(N)\times\mathbb{Z}$ can be deduced from the effective action
for the $(2+1)$d SPT phase coupled to a $PSU(N) \times \mathbb{Z}$ background field,
as discussed in~\cite{Append}.
As a result, we find that it is represented by a $\mathbb{Z}_N$ number,
$b_{\rho} \mod N$,
associated to an $SU(N)$ representation (rep.) $\rho$, 
where $b_{\rho}$ is the number of Young-tableau (YT) boxes in $\rho$.
Then, for a translationally invariant spin chain with an $SU(N)$ rep. $\lambda$ per site (unit cell), the above form of the mixed anomaly implies, from the matching condition (\ref{anomaly-matching}), its LSM index must be
$\mathcal{I}_{N} = b_\lambda \mod N$. 
For a more general $SU(N)$ model, the LSM index is given by the total number of YT boxes summed over all spins inside the unit cell:
\begin{align}
\label{LSM_index_general}
\mathcal{I}_{N} &= (\text{$\#$ of YT boxes per unit cell}) \mod N.
\end{align}
This can be argued in the following way.
Suppose there is a collection of $SU(N)$ spins with YT reps. $\{\lambda_i\}$ within a unit cell.
Then, we rearrange the system into a system of coupled spin chains, each of which has 
a $\lambda_i$-rep. spin per site.
Due to the additivity of anomaly factors and the definition~(\ref{anomaly-matching}), 
the LSM index of the whole system is just the sum of the associated indices of these spin chains, which equals the total number of YT boxes $\sum_{\lambda_i}b_{\lambda_i}$ per unit cell in the original model, namely, Eq.~(\ref{LSM_index_general}).
In Table~\ref{example} (the third column), we list the LSM indices for several $SU(N)$ models with given YT reps. per unit cell.
A system with a nonzero LSM index ($\mathcal{I}_N\neq 0\mod N$) must exhibit nontrivial low-energy behaviors in connection with the ingappability mentioned earlier, as we will elaborate in the following.

\textit{Ground-state degeneracy associated with a spontaneous broken translation symmetry ---} 
First let us assume that the system has a non-zero gap above the ground state(s).
In this case, a non-zero LSM index implies GSD, as in the case of the existing LSM-type theorems. Here we derive the degeneracy based on the mixed anomaly.
By considering a family of the LSM indices $p\mathcal{I}_N$ associated with lower translation symmetries $p\mathbb{Z}^\text{trans}\subseteq \mathbb{Z}^\text{trans}$ for $p\in\mathbb{N}$ of an $SU(N)$ spin model, we obtain a restriction on the GSD of any gapped phase of this model~\cite{Append}
\begin{align}
\label{GSD}
\mathrm{GSD}\in \frac{N}{\text{gcd}(\mathcal{I}_{N},N)}\mathbb{N}, 
\end{align}
{and the translation symmetry is spontaneously broken to at least $N/\text{gcd}(\mathcal{I}_N,N)$ of unit cells, realizable by exactly solvable models~\cite{Greiter:2007aa}.}
This indeed corresponds to the LSM theorem for the $SU(N)$ spin chain~\cite{Affleck:1986aa}, with an explicit statement on GSD.
In the first two rows of Table~\ref{example}, we list the exactly solvable $SU(3)$ trimer and {\bf 8}-VBS models -- analogs of $SU(2)$ dimer and AKLT models -- whose ground states break and respect a single-site translation symmetry, respectively. Their GSDs (shown in the fourth column) are consistent with the constraint (\ref{GSD}).

\textit{Constraint on critical WZW $SU(N)_k$ universality classes ---} 
Next we consider the other possibility, namely when the system is gapless.
While the usual LSM-type theorems do not give any further restriction in this case, the anomaly-based approach leads to constraints on the possible universality class of the gapless critical phase.
The most natural universality classes of a critical $SU(N)$ spin model is the $SU(N)$ WZW theories, with the global $SU(N)$ and conformal symmetries.
Their action is given as
 \begin{align}
 \label{WZW}
kI(g)&=\frac{k}{8\pi}\int_{M_2} dtdx\mathrm{Tr}\left(\partial_\mu g^{-1}\partial^\mu g\right)+k\Gamma_\text{WZ},  \nonumber\\
\Gamma_\text{WZ}&=\frac{1}{12\pi}\int_{B:\partial B=M_2}dtdxdy\text{Tr}(dgg^{-1})^3, 
\end{align}
where $k$ is an integer called level, $g$ is an $SU(N)$ matrix-valued field, and $\mathrm{Tr}$ is the conventional matrix trace.
The level $k$ characterizes the $SU(N)$ WZW theory, and we denote the $SU(N)$ WZW theory of level $k$ as $SU(N)_k$ WZW.
The Wess-Zumino term $\Gamma_\text{WZ}$ is defined as an extended integral onto an auxiliary manifold $B$ whose boundary is $M_2$, and a consistent CFT is independent on such extension. The lattice translation symmetry in the infrared becomes a discrete ``axial'' symmetry \cite{Affleck:1988aa} 
\begin{align}
\label{axial}
g\rightarrow e^{2\pi im/N}g, 
\end{align}
which forms a $\mathbb{Z}_n$ group with $n=N/\text{gcd}(m,N)$ and the integer $m$ defined modulo $N$.
(Note that any low-energy effective symmetry associated to the translation symmetry must be a subgroup of $\mathbb{Z}$.)
On the other hand, the on-site spin-rotation symmetry  corresponds to the vector $PSU(N)$ symmetry, a diagonal subgroup of a larger $[SU(N)_\mathrm{L}\times SU(N)_\mathrm{R}]/\mathbb{Z}_N$ symmetry, in the WZW theory.

As $PSU(N)\times \mathbb{Z}_n$ is exact as a global symmetry of the (quantum) WZW theory, there might be 't Hooft anomalies of it, including the mixed one between $PSU(N)$ and $\mathbb{Z}_n$ and the one for $\mathbb{Z}_n$ itself
\footnote{
There is no 't Hooft anomaly for $PSU(N)$ itself
}.
However, only the mixed anomaly is relevant to the physics characterized by the LSM index, while the the other one is an ``emergent'' anomaly 
which only characterizes ingappability against infinitesimal perturbations around criticality
\cite{Cho:2017aa, Metlitski-Thorngren17}
and is irrelevant to the situation we consider here.

One way to visualize the mixed anomaly in the WZW theory is to make use of the equivalence between $SU(N)_k$ WZW theory to $U(kN)_1/U(k)$ constrained Dirac fermion (CDF) theory
\cite{Naculich-Schnitzer90, Kiritsis-Niarchos10}
and then compute the $\mathbb{Z}_n$ axial anomaly in the CDF theory coupled to a background $PSU(N)$ gauge field. More explicitly, we show that~\cite{Append} 
there is a phase ambiguity of the WZW/CDF  partition function with a background $PSU(N)$ field under the $\mathbb{Z}_n$ transformation (\ref{axial}), which takes the form
$\exp\left({2\pi i km}\int_{M_2} w{(P)}/N \right)$,
where $P$ is the underlying $PSU(N)$ bundle (over $M_2$) and $w(P)\in H^2(M_2, \mathbb{Z}_N)\cong\mathbb{Z}_N$.
From this fact, we deduce that the mixed $PSU(N)-\mathbb{Z}_n$ anomaly for any $SU(N)_k$ WZW theory with $m$ defined by Eq.~(\ref{axial}) is characterized by a mod-$N$ integer 
\begin{align}
\label{mixed_anomaly}
km \mod N.
\end{align}
Note that the anomaly computed here is actually the mixed $PSU(N)-\mathbb{Z}$ anomaly discussed before, since the value of the mixed anomaly is unchanged when $\mathbb{Z}_n$ is extended to $\mathbb{Z}$ (by $n\mathbb{Z}$);
see similar discussions in 
\cite{Cho:2017aa, Metlitski-Thorngren17}.

Then, by the definition of the LSM index (\ref{anomaly-matching}) and also the way we represent it, we conclude that an $SU(N)$ model with an index $\mathcal{I}_{N}$ at the lattice scale, when flowing to some $SU(N)_k$ WZW CFT in the infrared, must obey 
\begin{align}
\label{matching_LSM_WZW}
\mathcal{I}_{N} = km \mod N.
\end{align}

\textit{Constraints on the $1+1$d $SU(N)$ spin models and symmetry-protected critical phases.---} Summarizing the above discussion, we obtain the following statement based on the matching condition, which includes the $SU(N)$ version of the LSM theorem:
\textit{
If a spin model with an exact $SU(N)$ spin-rotation and lattice translation symmetries has
a nontrivial LSM index $\mathcal{I}_N$~(\ref{LSM_index_general}), that is, the total umber of Young-tableau boxes per unit cell is not divisible by $N$,  
the system must either
\begin{itemize}
 \item have degenerate ground states below the gap, with the multiplicity~\eqref{GSD}, or
 \item have gapless excitations. If the low-energy EFT is given by an $SU(N)$ WZW theory, its level is constrained by the matching condition~\eqref{matching_LSM_WZW}.
\end{itemize}
}

The latter half of the statement implies an SPC classification of
the gapless critical systems with the global $SU(N)$ rotation and the
lattice translation symmetries.
That is, between the two fixed points
corresponding to $SU(N)_k$ and $SU(N)_{k'}$ WZW theories
with the representation of the translation symmetry~\eqref{axial} with
the factors $m$ and $m'$, a renormalization-group (RG) flow is only possible if
$km = k'm' \mod{N}$.

Our constraints are consistent with massless RG flows proposed in the literature~\cite{Schulz:1986aa,Ziman:1987aa,Lecheminant:2015aa,Lajko:2017aa}.
In Table~\ref{example}, we list several critical phases of $SU(N)$ spin systems with their numerically proposed IR CFTs in the fifth column. The mixed anomalies of these CFTs in the sixth column calculated by Eq.~(\ref{mixed_anomaly}) exactly match their LSM indices in the third column, consistent with our ``anomaly matching'' in Eq.~(\ref{matching_LSM_WZW}) and each CFT belongs to a certain SPC class of the underlying spin system. More specifically, the quantity $m$ of each IR CFT is calculated by the peak $\kappa=2\pi m/N$ of the structure factor defined as the Fourier transform of spin-spin correlations $C(r)\equiv\sum_{\alpha,\beta}\langle S^\alpha_{r,\beta} S^\beta_{0,\alpha}\rangle$~\cite{Dufour:2015aa}. In addition, perturbative half-integer-spin TB models are permitted to flow to $SU(2)_1$ WZW theory with $m=1$ since the CFT mixed anomaly matches their LSM indices, reproducing the $SU(2)$ SPC classification~\cite{Furuya:2017aa}.

\textit{Critical phases with higher symmetries ---} 
It is possible that an $SU(N)$ model can have a critical phase described by an $SU(N')_{k'}$ WZW CFT with $N'>N$. In this case, the critical theory is constrained, regarding the level $k'$ and the $\mathbb{Z}$ symmetry represented by $g\rightarrow e^{2\pi im'/N'}g$, by the $SU(N)$ LSM index $\mathcal{I}_N$ of the spin model through the following condition~\cite{Append}
\begin{align}
\label{matching_emergent}
\mathcal{I}_N\cdot\frac{N'}{\text{gcd}(N',k'm')}=0 \mod N.
\end{align}
As a special example, an $SU(2)$ spin chain with a half-integer spin per unit cell does not admit any critical phase described by the $SU(N')$ WZW CFT for any odd integer $N'$, which explains $SU(3)$ symmetries are only found in integer-spin models~\cite{Uimin:1970aa,Lai:1974aa,Sutherland:1975aa,Chen:2015aa}. 
Furthermore, if the $SU(N)$ spin model has an explicit symmetry enhancement (from $SU(N)$ to $SU(N')$) leading to the occurrence of a higher-symmetry critical phase, one can use a finer LSM index $\mathcal{I}_{N'}$ associated to the enlarged $SU(N')$ symmetry than the original index $\mathcal{I}_{N}$ to make further constraints on the possible critical theories.
For example, the $SU(2)$ spin-1 Uimin-Lai-Sutherland model~\cite{Uimin:1970aa,Lai:1974aa,Sutherland:1975aa}, which exhibits an explicit $SU(3)$ symmetry and can be expressed as an $SU(3)$ HAF model with fundamental representation per site (unit cell), has a nontrivial $SU(3)$ LSM index $\mathcal{I}_3=1\mod 3$ (while its $SU(2)$ LSM index $\mathcal{I}_2$ is trivial), consistent with the existence of the $SU(3)_1$ WZW critical theory (with $m'=1$) of this model.

\textit{Conclusions.---}
We proposed a topological ``LSM index'' to diagnose the ingappability of a generic $SU(N)$ spin system in $1+1$d with the spin-rotation and translation symmetries, based on the mixed 't Hooft anomaly.
It leads to a constraint on the GSD if the system is gapped (LSM theorem), or on the possible critical theory if the system is gapless.
Another implication is that $SU(N)$ WZW universality classes, which are SPC phases of the spin systems, fall into a $\mathbb{Z}_N$ classification.
Furthermore, the formalism can be applied to cases where a higher $SU(N')$ symmetry
emerges in an $SU(N)$-symmetric systems, to derive constraints on the possible phases
with the emergent symmetry.
We have verified that our results are consistent with several previous results listed in Table~\ref{example}.
Our approach is systematic and is not restricted to $SU(N)$. We believe that it will be useful to further explore SPC phases, in particular those with emergent symmetries.

\textit{Acknowledgements.---}
The authors thank Ian Affleck, Bo Han, Shinsei Ryu, Yuji Tachikawa, David Tong,
Keisuke Totsuka, and Kazuya Yonekura for useful discussions.
This research was supported in part by MEXT/JSPS KAKENHI Grant Numbers JP16K05469 and JP17H06462.
A part of the present work was also performed at Kavli Institute for Theoretical Physics, UC Santa Barbara, supported by the U. S. National Science Foundation under Grant No. NSF PHY-1748958.

\bibliography{bib}

\begin{thebibliography}{64}%
\makeatletter
\providecommand \@ifxundefined [1]{%
 \@ifx{#1\undefined}
}%
\providecommand \@ifnum [1]{%
 \ifnum #1\expandafter \@firstoftwo
 \else \expandafter \@secondoftwo
 \fi
}%
\providecommand \@ifx [1]{%
 \ifx #1\expandafter \@firstoftwo
 \else \expandafter \@secondoftwo
 \fi
}%
\providecommand \natexlab [1]{#1}%
\providecommand \enquote  [1]{``#1''}%
\providecommand \bibnamefont  [1]{#1}%
\providecommand \bibfnamefont [1]{#1}%
\providecommand \citenamefont [1]{#1}%
\providecommand \href@noop [0]{\@secondoftwo}%
\providecommand \href [0]{\begingroup \@sanitize@url \@href}%
\providecommand \@href[1]{\@@startlink{#1}\@@href}%
\providecommand \@@href[1]{\endgroup#1\@@endlink}%
\providecommand \@sanitize@url [0]{\catcode `\\12\catcode `\$12\catcode
  `\&12\catcode `\#12\catcode `\^12\catcode `\_12\catcode `\%12\relax}%
\providecommand \@@startlink[1]{}%
\providecommand \@@endlink[0]{}%
\providecommand \url  [0]{\begingroup\@sanitize@url \@url }%
\providecommand \@url [1]{\endgroup\@href {#1}{\urlprefix }}%
\providecommand \urlprefix  [0]{URL }%
\providecommand \Eprint [0]{\href }%
\providecommand \doibase [0]{http://dx.doi.org/}%
\providecommand \selectlanguage [0]{\@gobble}%
\providecommand \bibinfo  [0]{\@secondoftwo}%
\providecommand \bibfield  [0]{\@secondoftwo}%
\providecommand \translation [1]{[#1]}%
\providecommand \BibitemOpen [0]{}%
\providecommand \bibitemStop [0]{}%
\providecommand \bibitemNoStop [0]{.\EOS\space}%
\providecommand \EOS [0]{\spacefactor3000\relax}%
\providecommand \BibitemShut  [1]{\csname bibitem#1\endcsname}%
\let\auto@bib@innerbib\@empty
%</preamble>
\bibitem [{\citenamefont {Lieb}\ \emph {et~al.}(1961)\citenamefont {Lieb},
  \citenamefont {Schultz},\ and\ \citenamefont {Mattis}}]{Lieb:1961aa}%
  \BibitemOpen
  \bibfield  {author} {\bibinfo {author} {\bibfnamefont {E.}~\bibnamefont
  {Lieb}}, \bibinfo {author} {\bibfnamefont {T.}~\bibnamefont {Schultz}}, \
  and\ \bibinfo {author} {\bibfnamefont {D.}~\bibnamefont {Mattis}},\
  }\href@noop {} {\bibfield  {journal} {\bibinfo  {journal} {Ann. Phys.}\
  }\textbf {\bibinfo {volume} {16}},\ \bibinfo {pages} {407} (\bibinfo {year}
  {1961})}\BibitemShut {NoStop}%
\bibitem [{\citenamefont {Affleck}\ and\ \citenamefont
  {Lieb}(1986)}]{Affleck:1986aa}%
  \BibitemOpen
  \bibfield  {author} {\bibinfo {author} {\bibfnamefont {I.}~\bibnamefont
  {Affleck}}\ and\ \bibinfo {author} {\bibfnamefont {E.~H.}\ \bibnamefont
  {Lieb}},\ }\href@noop {} {\bibfield  {journal} {\bibinfo  {journal} {Lett.
  Math. Phys.}\ }\textbf {\bibinfo {volume} {12}},\ \bibinfo {pages} {57}
  (\bibinfo {year} {1986})}\BibitemShut {NoStop}%
\bibitem [{\citenamefont {Oshikawa}\ \emph {et~al.}(1997)\citenamefont
  {Oshikawa}, \citenamefont {Yamanaka},\ and\ \citenamefont
  {Affleck}}]{OYA1997}%
  \BibitemOpen
  \bibfield  {author} {\bibinfo {author} {\bibfnamefont {M.}~\bibnamefont
  {Oshikawa}}, \bibinfo {author} {\bibfnamefont {M.}~\bibnamefont {Yamanaka}},
  \ and\ \bibinfo {author} {\bibfnamefont {I.}~\bibnamefont {Affleck}},\ }\href
  {\doibase 10.1103/PhysRevLett.78.1984} {\bibfield  {journal} {\bibinfo
  {journal} {Phys. Rev. Lett.}\ }\textbf {\bibinfo {volume} {78}},\ \bibinfo
  {pages} {1984} (\bibinfo {year} {1997})}\BibitemShut {NoStop}%
\bibitem [{\citenamefont {Oshikawa}(2000)}]{Oshikawa:2000aa}%
  \BibitemOpen
  \bibfield  {author} {\bibinfo {author} {\bibfnamefont {M.}~\bibnamefont
  {Oshikawa}},\ }\href {https://link.aps.org/doi/10.1103/PhysRevLett.84.1535}
  {\bibfield  {journal} {\bibinfo  {journal} {Phys. Rev. Lett.}\ }\textbf
  {\bibinfo {volume} {84}},\ \bibinfo {pages} {1535} (\bibinfo {year}
  {2000})}\BibitemShut {NoStop}%
\bibitem [{\citenamefont {Hastings}(2004)}]{Hastings:2004ab}%
  \BibitemOpen
  \bibfield  {author} {\bibinfo {author} {\bibfnamefont {M.~B.}\ \bibnamefont
  {Hastings}},\ }\href {https://link.aps.org/doi/10.1103/PhysRevB.69.104431}
  {\bibfield  {journal} {\bibinfo  {journal} {Phys. Rev. B}\ }\textbf {\bibinfo
  {volume} {69}},\ \bibinfo {pages} {104431} (\bibinfo {year}
  {2004})}\BibitemShut {NoStop}%
\bibitem [{\citenamefont {Alcaraz}\ and\ \citenamefont
  {Martins}(1989)}]{Alcaraz:1989aa}%
  \BibitemOpen
  \bibfield  {author} {\bibinfo {author} {\bibfnamefont {F.~C.}\ \bibnamefont
  {Alcaraz}}\ and\ \bibinfo {author} {\bibfnamefont {M.~J.}\ \bibnamefont
  {Martins}},\ }\href@noop {} {\bibfield  {journal} {\bibinfo  {journal} {J.
  Phys. A: Math. Theor.}\ }\textbf {\bibinfo {volume} {22}},\ \bibinfo {pages}
  {L865} (\bibinfo {year} {1989})}\BibitemShut {NoStop}%
\bibitem [{\citenamefont {F{\"u}hringer}\ \emph {et~al.}(2008)\citenamefont
  {F{\"u}hringer}, \citenamefont {Rachel}, \citenamefont {Thomale},
  \citenamefont {Greiter},\ and\ \citenamefont
  {Schmitteckert}}]{Fuhringer:2008aa}%
  \BibitemOpen
  \bibfield  {author} {\bibinfo {author} {\bibfnamefont {M.}~\bibnamefont
  {F{\"u}hringer}}, \bibinfo {author} {\bibfnamefont {S.}~\bibnamefont
  {Rachel}}, \bibinfo {author} {\bibfnamefont {R.}~\bibnamefont {Thomale}},
  \bibinfo {author} {\bibfnamefont {M.}~\bibnamefont {Greiter}}, \ and\
  \bibinfo {author} {\bibfnamefont {P.}~\bibnamefont {Schmitteckert}},\
  }\href@noop {} {\bibfield  {journal} {\bibinfo  {journal} {Ann. Phys.
  (Berlin)}\ }\textbf {\bibinfo {volume} {17}},\ \bibinfo {pages} {922}
  (\bibinfo {year} {2008})}\BibitemShut {NoStop}%
\bibitem [{\citenamefont {Matsumoto}\ \emph {et~al.}(2011)\citenamefont
  {Matsumoto}, \citenamefont {Nakatsuji}, \citenamefont {Kuga}, \citenamefont
  {Karaki}, \citenamefont {Horie}, \citenamefont {Shimura}, \citenamefont
  {Sakakibara}, \citenamefont {Nevidomskyy},\ and\ \citenamefont
  {Coleman}}]{Matsumoto-YbAlB4_Science2011}%
  \BibitemOpen
  \bibfield  {author} {\bibinfo {author} {\bibfnamefont {Y.}~\bibnamefont
  {Matsumoto}}, \bibinfo {author} {\bibfnamefont {S.}~\bibnamefont
  {Nakatsuji}}, \bibinfo {author} {\bibfnamefont {K.}~\bibnamefont {Kuga}},
  \bibinfo {author} {\bibfnamefont {Y.}~\bibnamefont {Karaki}}, \bibinfo
  {author} {\bibfnamefont {N.}~\bibnamefont {Horie}}, \bibinfo {author}
  {\bibfnamefont {Y.}~\bibnamefont {Shimura}}, \bibinfo {author} {\bibfnamefont
  {T.}~\bibnamefont {Sakakibara}}, \bibinfo {author} {\bibfnamefont {A.~H.}\
  \bibnamefont {Nevidomskyy}}, \ and\ \bibinfo {author} {\bibfnamefont
  {P.}~\bibnamefont {Coleman}},\ }\href {\doibase 10.1126/science.1197531}
  {\bibfield  {journal} {\bibinfo  {journal} {Science}\ }\textbf {\bibinfo
  {volume} {331}},\ \bibinfo {pages} {316} (\bibinfo {year}
  {2011})}\BibitemShut {NoStop}%
\bibitem [{\citenamefont {Furuya}\ and\ \citenamefont
  {Oshikawa}(2017)}]{Furuya:2017aa}%
  \BibitemOpen
  \bibfield  {author} {\bibinfo {author} {\bibfnamefont {S.~C.}\ \bibnamefont
  {Furuya}}\ and\ \bibinfo {author} {\bibfnamefont {M.}~\bibnamefont
  {Oshikawa}},\ }\href
  {https://link.aps.org/doi/10.1103/PhysRevLett.118.021601} {\bibfield
  {journal} {\bibinfo  {journal} {Phys. Rev. Lett.}\ }\textbf {\bibinfo
  {volume} {118}},\ \bibinfo {pages} {021601} (\bibinfo {year}
  {2017})}\BibitemShut {NoStop}%
\bibitem [{\citenamefont {Wu}\ \emph {et~al.}(2003)\citenamefont {Wu},
  \citenamefont {Hu},\ and\ \citenamefont {Zhang}}]{Wu:2003aa}%
  \BibitemOpen
  \bibfield  {author} {\bibinfo {author} {\bibfnamefont {C.}~\bibnamefont
  {Wu}}, \bibinfo {author} {\bibfnamefont {J.-P.}\ \bibnamefont {Hu}}, \ and\
  \bibinfo {author} {\bibfnamefont {S.-C.}\ \bibnamefont {Zhang}},\ }\href
  {https://link.aps.org/doi/10.1103/PhysRevLett.91.186402} {\bibfield
  {journal} {\bibinfo  {journal} {Phys. Rev. Lett.}\ }\textbf {\bibinfo
  {volume} {91}},\ \bibinfo {pages} {186402} (\bibinfo {year}
  {2003})}\BibitemShut {NoStop}%
\bibitem [{\citenamefont {Honerkamp}\ and\ \citenamefont
  {Hofstetter}(2004)}]{Honerkamp:2004aa}%
  \BibitemOpen
  \bibfield  {author} {\bibinfo {author} {\bibfnamefont {C.}~\bibnamefont
  {Honerkamp}}\ and\ \bibinfo {author} {\bibfnamefont {W.}~\bibnamefont
  {Hofstetter}},\ }\href
  {https://link.aps.org/doi/10.1103/PhysRevLett.92.170403} {\bibfield
  {journal} {\bibinfo  {journal} {Phys. Rev. Lett.}\ }\textbf {\bibinfo
  {volume} {92}},\ \bibinfo {pages} {170403} (\bibinfo {year}
  {2004})}\BibitemShut {NoStop}%
\bibitem [{\citenamefont {Cazalilla}\ \emph {et~al.}(2009)\citenamefont
  {Cazalilla}, \citenamefont {Ho},\ and\ \citenamefont
  {Ueda}}]{Cazalilla:2009aa}%
  \BibitemOpen
  \bibfield  {author} {\bibinfo {author} {\bibfnamefont {M.~A.}\ \bibnamefont
  {Cazalilla}}, \bibinfo {author} {\bibfnamefont {A.}~\bibnamefont {Ho}}, \
  and\ \bibinfo {author} {\bibfnamefont {M.}~\bibnamefont {Ueda}},\ }\href@noop
  {} {\bibfield  {journal} {\bibinfo  {journal} {New J. Phys.}\ }\textbf
  {\bibinfo {volume} {11}},\ \bibinfo {pages} {103033 1367} (\bibinfo {year}
  {2009})}\BibitemShut {NoStop}%
\bibitem [{\citenamefont {Gorshkov}\ \emph {et~al.}(2010)\citenamefont
  {Gorshkov}, \citenamefont {Hermele}, \citenamefont {Gurarie}, \citenamefont
  {Xu}, \citenamefont {Julienne}, \citenamefont {Ye}, \citenamefont {Zoller},
  \citenamefont {Demler}, \citenamefont {Lukin},\ and\ \citenamefont
  {Rey}}]{Gorshkov:2010aa}%
  \BibitemOpen
  \bibfield  {author} {\bibinfo {author} {\bibfnamefont {A.~V.}\ \bibnamefont
  {Gorshkov}}, \bibinfo {author} {\bibfnamefont {M.}~\bibnamefont {Hermele}},
  \bibinfo {author} {\bibfnamefont {V.}~\bibnamefont {Gurarie}}, \bibinfo
  {author} {\bibfnamefont {C.}~\bibnamefont {Xu}}, \bibinfo {author}
  {\bibfnamefont {P.~S.}\ \bibnamefont {Julienne}}, \bibinfo {author}
  {\bibfnamefont {J.}~\bibnamefont {Ye}}, \bibinfo {author} {\bibfnamefont
  {P.}~\bibnamefont {Zoller}}, \bibinfo {author} {\bibfnamefont
  {E.}~\bibnamefont {Demler}}, \bibinfo {author} {\bibfnamefont {M.~D.}\
  \bibnamefont {Lukin}}, \ and\ \bibinfo {author} {\bibfnamefont
  {A.}~\bibnamefont {Rey}},\ }\href@noop {} {\bibfield  {journal} {\bibinfo
  {journal} {Nature physics}\ }\textbf {\bibinfo {volume} {6}},\ \bibinfo
  {pages} {289} (\bibinfo {year} {2010})}\BibitemShut {NoStop}%
\bibitem [{\citenamefont {Taie}\ \emph {et~al.}(2012)\citenamefont {Taie},
  \citenamefont {Yamazaki}, \citenamefont {Sugawa},\ and\ \citenamefont
  {Takahashi}}]{Taie:2012aa}%
  \BibitemOpen
  \bibfield  {author} {\bibinfo {author} {\bibfnamefont {S.}~\bibnamefont
  {Taie}}, \bibinfo {author} {\bibfnamefont {R.}~\bibnamefont {Yamazaki}},
  \bibinfo {author} {\bibfnamefont {S.}~\bibnamefont {Sugawa}}, \ and\ \bibinfo
  {author} {\bibfnamefont {Y.}~\bibnamefont {Takahashi}},\ }\href@noop {}
  {\bibfield  {journal} {\bibinfo  {journal} {Nature Physics}\ }\textbf
  {\bibinfo {volume} {8}},\ \bibinfo {pages} {825} (\bibinfo {year}
  {2012})}\BibitemShut {NoStop}%
\bibitem [{\citenamefont {Pagano}\ \emph {et~al.}(2014)\citenamefont {Pagano},
  \citenamefont {Mancini}, \citenamefont {Cappellini}, \citenamefont
  {Lombardi}, \citenamefont {Sch{\"a}fer}, \citenamefont {Hu}, \citenamefont
  {Liu}, \citenamefont {Catani}, \citenamefont {Sias},\ and\ \citenamefont
  {Inguscio}}]{Pagano:2014aa}%
  \BibitemOpen
  \bibfield  {author} {\bibinfo {author} {\bibfnamefont {G.}~\bibnamefont
  {Pagano}}, \bibinfo {author} {\bibfnamefont {M.}~\bibnamefont {Mancini}},
  \bibinfo {author} {\bibfnamefont {G.}~\bibnamefont {Cappellini}}, \bibinfo
  {author} {\bibfnamefont {P.}~\bibnamefont {Lombardi}}, \bibinfo {author}
  {\bibfnamefont {F.}~\bibnamefont {Sch{\"a}fer}}, \bibinfo {author}
  {\bibfnamefont {H.}~\bibnamefont {Hu}}, \bibinfo {author} {\bibfnamefont
  {X.-J.}\ \bibnamefont {Liu}}, \bibinfo {author} {\bibfnamefont
  {J.}~\bibnamefont {Catani}}, \bibinfo {author} {\bibfnamefont
  {C.}~\bibnamefont {Sias}}, \ and\ \bibinfo {author} {\bibfnamefont
  {M.}~\bibnamefont {Inguscio}},\ }\href@noop {} {\bibfield  {journal}
  {\bibinfo  {journal} {Nature Physics}\ }\textbf {\bibinfo {volume} {10}},\
  \bibinfo {pages} {198} (\bibinfo {year} {2014})}\BibitemShut {NoStop}%
\bibitem [{\citenamefont {Scazza}\ \emph {et~al.}(2014)\citenamefont {Scazza},
  \citenamefont {Hofrichter}, \citenamefont {H{\"o}fer}, \citenamefont
  {De~Groot}, \citenamefont {Bloch},\ and\ \citenamefont
  {F{\"o}lling}}]{Scazza:2014aa}%
  \BibitemOpen
  \bibfield  {author} {\bibinfo {author} {\bibfnamefont {F.}~\bibnamefont
  {Scazza}}, \bibinfo {author} {\bibfnamefont {C.}~\bibnamefont {Hofrichter}},
  \bibinfo {author} {\bibfnamefont {M.}~\bibnamefont {H{\"o}fer}}, \bibinfo
  {author} {\bibfnamefont {P.}~\bibnamefont {De~Groot}}, \bibinfo {author}
  {\bibfnamefont {I.}~\bibnamefont {Bloch}}, \ and\ \bibinfo {author}
  {\bibfnamefont {S.}~\bibnamefont {F{\"o}lling}},\ }\href@noop {} {\bibfield
  {journal} {\bibinfo  {journal} {Nature Physics}\ }\textbf {\bibinfo {volume}
  {10}},\ \bibinfo {pages} {779} (\bibinfo {year} {2014})}\BibitemShut
  {NoStop}%
\bibitem [{\citenamefont {Zhang}\ \emph {et~al.}(2014)\citenamefont {Zhang},
  \citenamefont {Bishof}, \citenamefont {Bromley}, \citenamefont {Kraus},
  \citenamefont {Safronova}, \citenamefont {Zoller}, \citenamefont {Rey},\ and\
  \citenamefont {Ye}}]{Zhang:2014aa}%
  \BibitemOpen
  \bibfield  {author} {\bibinfo {author} {\bibfnamefont {X.}~\bibnamefont
  {Zhang}}, \bibinfo {author} {\bibfnamefont {M.}~\bibnamefont {Bishof}},
  \bibinfo {author} {\bibfnamefont {S.}~\bibnamefont {Bromley}}, \bibinfo
  {author} {\bibfnamefont {C.}~\bibnamefont {Kraus}}, \bibinfo {author}
  {\bibfnamefont {M.}~\bibnamefont {Safronova}}, \bibinfo {author}
  {\bibfnamefont {P.}~\bibnamefont {Zoller}}, \bibinfo {author} {\bibfnamefont
  {A.~M.}\ \bibnamefont {Rey}}, \ and\ \bibinfo {author} {\bibfnamefont
  {J.}~\bibnamefont {Ye}},\ }\href@noop {} {\bibfield  {journal} {\bibinfo
  {journal} {science}\ }\textbf {\bibinfo {volume} {345}},\ \bibinfo {pages}
  {1467} (\bibinfo {year} {2014})}\BibitemShut {NoStop}%
\bibitem [{\citenamefont {{Yamada}}\ \emph {et~al.}(2017)\citenamefont
  {{Yamada}}, \citenamefont {{Oshikawa}},\ and\ \citenamefont
  {{Jackeli}}}]{YamadaSU4}%
  \BibitemOpen
  \bibfield  {author} {\bibinfo {author} {\bibfnamefont {M.~G.}\ \bibnamefont
  {{Yamada}}}, \bibinfo {author} {\bibfnamefont {M.}~\bibnamefont
  {{Oshikawa}}}, \ and\ \bibinfo {author} {\bibfnamefont {G.}~\bibnamefont
  {{Jackeli}}},\ }\href@noop {} {\bibfield  {journal} {\bibinfo  {journal}
  {ArXiv e-prints}\ } (\bibinfo {year} {2017})},\ \Eprint
  {http://arxiv.org/abs/1709.05252} {arXiv:1709.05252 [cond-mat.str-el]}
  \BibitemShut {NoStop}%
\bibitem [{\citenamefont {Uimin}(1970)}]{Uimin:1970aa}%
  \BibitemOpen
  \bibfield  {author} {\bibinfo {author} {\bibfnamefont {G.}~\bibnamefont
  {Uimin}},\ }\href@noop {} {\bibfield  {journal} {\bibinfo  {journal} {JETP
  Lett.}\ }\textbf {\bibinfo {volume} {12}},\ \bibinfo {pages} {225} (\bibinfo
  {year} {1970})}\BibitemShut {NoStop}%
\bibitem [{\citenamefont {Lai}(1974)}]{Lai:1974aa}%
  \BibitemOpen
  \bibfield  {author} {\bibinfo {author} {\bibfnamefont {C.}~\bibnamefont
  {Lai}},\ }\href@noop {} {\bibfield  {journal} {\bibinfo  {journal} {J. Math.
  Phys.}\ }\textbf {\bibinfo {volume} {15}},\ \bibinfo {pages} {1675} (\bibinfo
  {year} {1974})}\BibitemShut {NoStop}%
\bibitem [{\citenamefont {Sutherland}(1975)}]{Sutherland:1975aa}%
  \BibitemOpen
  \bibfield  {author} {\bibinfo {author} {\bibfnamefont {B.}~\bibnamefont
  {Sutherland}},\ }\href {\doibase 10.1103/PhysRevB.12.3795} {\bibfield
  {journal} {\bibinfo  {journal} {Phys. Rev. B}\ }\textbf {\bibinfo {volume}
  {12}},\ \bibinfo {pages} {3795} (\bibinfo {year} {1975})}\BibitemShut
  {NoStop}%
\bibitem [{\citenamefont {Chen}\ \emph {et~al.}(2015)\citenamefont {Chen},
  \citenamefont {Xue}, \citenamefont {McCulloch}, \citenamefont {Chung},
  \citenamefont {Huang},\ and\ \citenamefont {Yip}}]{Chen:2015aa}%
  \BibitemOpen
  \bibfield  {author} {\bibinfo {author} {\bibfnamefont {P.}~\bibnamefont
  {Chen}}, \bibinfo {author} {\bibfnamefont {Z.-L.}\ \bibnamefont {Xue}},
  \bibinfo {author} {\bibfnamefont {I.}~\bibnamefont {McCulloch}}, \bibinfo
  {author} {\bibfnamefont {M.-C.}\ \bibnamefont {Chung}}, \bibinfo {author}
  {\bibfnamefont {C.-C.}\ \bibnamefont {Huang}}, \ and\ \bibinfo {author}
  {\bibfnamefont {S.-K.}\ \bibnamefont {Yip}},\ }\href@noop {} {\bibfield
  {journal} {\bibinfo  {journal} {Phys. Rev. Lett.}\ }\textbf {\bibinfo
  {volume} {114}},\ \bibinfo {pages} {145301} (\bibinfo {year}
  {2015})}\BibitemShut {NoStop}%
\bibitem [{\citenamefont {'t~Hooft}\ \emph {et~al.}(1980)\citenamefont
  {'t~Hooft}, \citenamefont {Itzykson}, \citenamefont {Jaffe}, \citenamefont
  {Lehmann}, \citenamefont {Mitter}, \citenamefont {Singer},\ and\
  \citenamefont {Stora}}]{tHooft:1980xss}%
  \BibitemOpen
  \bibfield  {author} {\bibinfo {author} {\bibfnamefont {G.}~\bibnamefont
  {'t~Hooft}}, \bibinfo {author} {\bibfnamefont {C.}~\bibnamefont {Itzykson}},
  \bibinfo {author} {\bibfnamefont {A.}~\bibnamefont {Jaffe}}, \bibinfo
  {author} {\bibfnamefont {H.}~\bibnamefont {Lehmann}}, \bibinfo {author}
  {\bibfnamefont {P.~K.}\ \bibnamefont {Mitter}}, \bibinfo {author}
  {\bibfnamefont {I.~M.}\ \bibnamefont {Singer}}, \ and\ \bibinfo {author}
  {\bibfnamefont {R.}~\bibnamefont {Stora}},\ }\href {\doibase
  10.1007/978-1-4684-7571-5} {\bibfield  {journal} {\bibinfo  {journal} {NATO
  Sci. Ser. B}\ }\textbf {\bibinfo {volume} {59}},\ \bibinfo {pages} {pp.1}
  (\bibinfo {year} {1980})}\BibitemShut {NoStop}%
%%CITATION = INSPIRE-161902;%%
\bibitem [{\citenamefont {Benini}\ \emph {et~al.}(2017)\citenamefont {Benini},
  \citenamefont {Hsin},\ and\ \citenamefont {Seiberg}}]{Benini:2017aa}%
  \BibitemOpen
  \bibfield  {author} {\bibinfo {author} {\bibfnamefont {F.}~\bibnamefont
  {Benini}}, \bibinfo {author} {\bibfnamefont {P.-S.}\ \bibnamefont {Hsin}}, \
  and\ \bibinfo {author} {\bibfnamefont {N.}~\bibnamefont {Seiberg}},\ }\href
  {\doibase 10.1007/JHEP04(2017)135} {\bibfield  {journal} {\bibinfo  {journal}
  {JHEP}\ }\textbf {\bibinfo {volume} {2017}},\ \bibinfo {pages} {135}
  (\bibinfo {year} {2017})}\BibitemShut {NoStop}%
\bibitem [{\citenamefont {Gaiotto}\ \emph {et~al.}(2017)\citenamefont
  {Gaiotto}, \citenamefont {Kapustin}, \citenamefont {Komargodski},\ and\
  \citenamefont {Seiberg}}]{Gaiotto:2017aa}%
  \BibitemOpen
  \bibfield  {author} {\bibinfo {author} {\bibfnamefont {D.}~\bibnamefont
  {Gaiotto}}, \bibinfo {author} {\bibfnamefont {A.}~\bibnamefont {Kapustin}},
  \bibinfo {author} {\bibfnamefont {Z.}~\bibnamefont {Komargodski}}, \ and\
  \bibinfo {author} {\bibfnamefont {N.}~\bibnamefont {Seiberg}},\ }\href
  {\doibase 10.1007/JHEP05(2017)091} {\bibfield  {journal} {\bibinfo  {journal}
  {JHEP}\ }\textbf {\bibinfo {volume} {2017}},\ \bibinfo {pages} {91} (\bibinfo
  {year} {2017})}\BibitemShut {NoStop}%
\bibitem [{\citenamefont {Tanizaki}\ and\ \citenamefont
  {Kikuchi}(2017)}]{Tanizaki:2017aa}%
  \BibitemOpen
  \bibfield  {author} {\bibinfo {author} {\bibfnamefont {Y.}~\bibnamefont
  {Tanizaki}}\ and\ \bibinfo {author} {\bibfnamefont {Y.}~\bibnamefont
  {Kikuchi}},\ }\href {\doibase 10.1007/JHEP06(2017)102} {\bibfield  {journal}
  {\bibinfo  {journal} {JHEP}\ }\textbf {\bibinfo {volume} {2017}},\ \bibinfo
  {pages} {102} (\bibinfo {year} {2017})}\BibitemShut {NoStop}%
\bibitem [{\citenamefont {Komargodski}\ \emph {et~al.}(2017)\citenamefont
  {Komargodski}, \citenamefont {Sharon}, \citenamefont {Thorngren},\ and\
  \citenamefont {Zhou}}]{Komargodski:2017aa}%
  \BibitemOpen
  \bibfield  {author} {\bibinfo {author} {\bibfnamefont {Z.}~\bibnamefont
  {Komargodski}}, \bibinfo {author} {\bibfnamefont {A.}~\bibnamefont {Sharon}},
  \bibinfo {author} {\bibfnamefont {R.}~\bibnamefont {Thorngren}}, \ and\
  \bibinfo {author} {\bibfnamefont {X.}~\bibnamefont {Zhou}},\ }\href@noop {}
  {\bibfield  {journal} {\bibinfo  {journal} {arXiv preprint,
  arXiv:1705.04786}\ } (\bibinfo {year} {2017})}\BibitemShut {NoStop}%
\bibitem [{\citenamefont {Komargodski}\ \emph {et~al.}(2018)\citenamefont
  {Komargodski}, \citenamefont {Sulejmanpasic},\ and\ \citenamefont
  {{\"U}nsal}}]{Komargodski:2018aa}%
  \BibitemOpen
  \bibfield  {author} {\bibinfo {author} {\bibfnamefont {Z.}~\bibnamefont
  {Komargodski}}, \bibinfo {author} {\bibfnamefont {T.}~\bibnamefont
  {Sulejmanpasic}}, \ and\ \bibinfo {author} {\bibfnamefont {M.}~\bibnamefont
  {{\"U}nsal}},\ }\href {\doibase 10.1103/PhysRevB.97.054418} {\bibfield
  {journal} {\bibinfo  {journal} {Phys. Rev. B}\ }\textbf {\bibinfo {volume}
  {97}},\ \bibinfo {pages} {054418} (\bibinfo {year} {2018})}\BibitemShut
  {NoStop}%
\bibitem [{\citenamefont {Shimizu}\ and\ \citenamefont
  {Yonekura}(2017)}]{Shimizu:2017aa}%
  \BibitemOpen
  \bibfield  {author} {\bibinfo {author} {\bibfnamefont {H.}~\bibnamefont
  {Shimizu}}\ and\ \bibinfo {author} {\bibfnamefont {K.}~\bibnamefont
  {Yonekura}},\ }\href@noop {} {\bibfield  {journal} {\bibinfo  {journal}
  {arXiv preprint, arXiv:1706.06104}\ } (\bibinfo {year} {2017})}\BibitemShut
  {NoStop}%
\bibitem [{Note1()}]{Note1}%
  \BibitemOpen
  \bibinfo {note} {For example, when the system is in a gapless phase, there
  are both left- and right-moving soft (chiral) modes present at low energy,
  and the $PSU(N)$ and $\protect \mathbb {Z}$ symmetries would act on these
  modes as a ``vector'' and an ``axial'' symmetries, respectively. In this
  situation, there is a potential chiral/axial anomaly in the low-energy field
  theory, similar to the case of the $(1+1)$d Dirac theory with both a vector
  and an axial $U(1)$ symmetries.}\BibitemShut {Stop}%
\bibitem [{\citenamefont {Cheng}\ \emph {et~al.}(2016)\citenamefont {Cheng},
  \citenamefont {Zaletel}, \citenamefont {Barkeshli}, \citenamefont
  {Vishwanath},\ and\ \citenamefont {Bonderson}}]{Cheng:2016aa}%
  \BibitemOpen
  \bibfield  {author} {\bibinfo {author} {\bibfnamefont {M.}~\bibnamefont
  {Cheng}}, \bibinfo {author} {\bibfnamefont {M.}~\bibnamefont {Zaletel}},
  \bibinfo {author} {\bibfnamefont {M.}~\bibnamefont {Barkeshli}}, \bibinfo
  {author} {\bibfnamefont {A.}~\bibnamefont {Vishwanath}}, \ and\ \bibinfo
  {author} {\bibfnamefont {P.}~\bibnamefont {Bonderson}},\ }\href {\doibase
  10.1103/PhysRevX.6.041068} {\bibfield  {journal} {\bibinfo  {journal} {Phys.
  Rev. X}\ }\textbf {\bibinfo {volume} {6}},\ \bibinfo {pages} {041068}
  (\bibinfo {year} {2016})}\BibitemShut {NoStop}%
\bibitem [{\citenamefont {Jian}\ \emph {et~al.}(2018)\citenamefont {Jian},
  \citenamefont {Bi},\ and\ \citenamefont {Xu}}]{Jian:2018aa}%
  \BibitemOpen
  \bibfield  {author} {\bibinfo {author} {\bibfnamefont {C.-M.}\ \bibnamefont
  {Jian}}, \bibinfo {author} {\bibfnamefont {Z.}~\bibnamefont {Bi}}, \ and\
  \bibinfo {author} {\bibfnamefont {C.}~\bibnamefont {Xu}},\ }\href {\doibase
  10.1103/PhysRevB.97.054412} {\bibfield  {journal} {\bibinfo  {journal} {Phys.
  Rev. B}\ }\textbf {\bibinfo {volume} {97}},\ \bibinfo {pages} {054412}
  (\bibinfo {year} {2018})}\BibitemShut {NoStop}%
\bibitem [{\citenamefont {Cho}\ \emph {et~al.}(2017)\citenamefont {Cho},
  \citenamefont {Hsieh},\ and\ \citenamefont {Ryu}}]{Cho:2017aa}%
  \BibitemOpen
  \bibfield  {author} {\bibinfo {author} {\bibfnamefont {G.~Y.}\ \bibnamefont
  {Cho}}, \bibinfo {author} {\bibfnamefont {C.-T.}\ \bibnamefont {Hsieh}}, \
  and\ \bibinfo {author} {\bibfnamefont {S.}~\bibnamefont {Ryu}},\ }\href
  {https://link.aps.org/doi/10.1103/PhysRevB.96.195105} {\bibfield  {journal}
  {\bibinfo  {journal} {Phys. Rev. B}\ }\textbf {\bibinfo {volume} {96}},\
  \bibinfo {pages} {195105} (\bibinfo {year} {2017})}\BibitemShut {NoStop}%
\bibitem [{\citenamefont {{ Metlitski}}\ and\ \citenamefont
  {{Thorngren}}(2017)}]{Metlitski-Thorngren17}%
  \BibitemOpen
  \bibfield  {author} {\bibinfo {author} {\bibfnamefont {M.~A.}\ \bibnamefont
  {{ Metlitski}}}\ and\ \bibinfo {author} {\bibfnamefont {R.}~\bibnamefont
  {{Thorngren}}},\ }\href@noop {} {\bibfield  {journal} {\bibinfo  {journal}
  {arXiv e-prints}\ } (\bibinfo {year} {2017})},\ \Eprint
  {http://arxiv.org/abs/1707.07686} {arXiv:1707.07686 [cond-mat.str-el]}
  \BibitemShut {NoStop}%
\bibitem [{\citenamefont {Cheng}(2018)}]{Cheng:2018aa}%
  \BibitemOpen
  \bibfield  {author} {\bibinfo {author} {\bibfnamefont {M.}~\bibnamefont
  {Cheng}},\ }\href@noop {} {\bibfield  {journal} {\bibinfo  {journal} {arXiv
  preprint arXiv:1804.10122}\ } (\bibinfo {year} {2018})}\BibitemShut {NoStop}%
\bibitem [{Note2()}]{Note2}%
  \BibitemOpen
  \bibinfo {note} {The $PSU(N)$-broken gapless phase is actually the
  ferromagnetic phase, which exists only at zero temperature and does not
  survive at finite temperature, according to the Mermin-Wagner
  Theorem.}\BibitemShut {Stop}%
\bibitem [{\citenamefont {Ryu}\ and\ \citenamefont {Zhang}(2012)}]{Ryu:2012ac}%
  \BibitemOpen
  \bibfield  {author} {\bibinfo {author} {\bibfnamefont {S.}~\bibnamefont
  {Ryu}}\ and\ \bibinfo {author} {\bibfnamefont {S.-C.}\ \bibnamefont
  {Zhang}},\ }\href {\doibase 10.1103/PhysRevB.85.245132} {\bibfield  {journal}
  {\bibinfo  {journal} {Phys. Rev. B}\ }\textbf {\bibinfo {volume} {85}},\
  \bibinfo {pages} {245132} (\bibinfo {year} {2012})}\BibitemShut {NoStop}%
\bibitem [{\citenamefont {Wen}(2013)}]{Wen:2013aa}%
  \BibitemOpen
  \bibfield  {author} {\bibinfo {author} {\bibfnamefont {X.-G.}\ \bibnamefont
  {Wen}},\ }\href {\doibase 10.1103/PhysRevD.88.045013} {\bibfield  {journal}
  {\bibinfo  {journal} {Phys. Rev. D}\ }\textbf {\bibinfo {volume} {88}},\
  \bibinfo {pages} {045013} (\bibinfo {year} {2013})}\BibitemShut {NoStop}%
\bibitem [{\citenamefont {Wang}\ and\ \citenamefont {Wen}(2013)}]{Wang:2013aa}%
  \BibitemOpen
  \bibfield  {author} {\bibinfo {author} {\bibfnamefont {J.}~\bibnamefont
  {Wang}}\ and\ \bibinfo {author} {\bibfnamefont {X.-G.}\ \bibnamefont {Wen}},\
  }\href@noop {} {\bibfield  {journal} {\bibinfo  {journal} {arXiv preprint
  arXiv:1307.7480}\ } (\bibinfo {year} {2013})}\BibitemShut {NoStop}%
\bibitem [{\citenamefont {Hsieh}\ \emph {et~al.}(2014)\citenamefont {Hsieh},
  \citenamefont {Sule}, \citenamefont {Cho}, \citenamefont {Ryu},\ and\
  \citenamefont {Leigh}}]{Hsieh:2014aa}%
  \BibitemOpen
  \bibfield  {author} {\bibinfo {author} {\bibfnamefont {C.-T.}\ \bibnamefont
  {Hsieh}}, \bibinfo {author} {\bibfnamefont {O.~M.}\ \bibnamefont {Sule}},
  \bibinfo {author} {\bibfnamefont {G.~Y.}\ \bibnamefont {Cho}}, \bibinfo
  {author} {\bibfnamefont {S.}~\bibnamefont {Ryu}}, \ and\ \bibinfo {author}
  {\bibfnamefont {R.~G.}\ \bibnamefont {Leigh}},\ }\href
  {https://link.aps.org/doi/10.1103/PhysRevB.90.165134} {\bibfield  {journal}
  {\bibinfo  {journal} {Phys. Rev. B}\ }\textbf {\bibinfo {volume} {90}},\
  \bibinfo {pages} {165134} (\bibinfo {year} {2014})}\BibitemShut {NoStop}%
\bibitem [{\citenamefont {Hsieh}\ \emph {et~al.}(2016)\citenamefont {Hsieh},
  \citenamefont {Cho},\ and\ \citenamefont {Ryu}}]{Hsieh:2016aa}%
  \BibitemOpen
  \bibfield  {author} {\bibinfo {author} {\bibfnamefont {C.-T.}\ \bibnamefont
  {Hsieh}}, \bibinfo {author} {\bibfnamefont {G.~Y.}\ \bibnamefont {Cho}}, \
  and\ \bibinfo {author} {\bibfnamefont {S.}~\bibnamefont {Ryu}},\ }\href
  {https://link.aps.org/doi/10.1103/PhysRevB.93.075135} {\bibfield  {journal}
  {\bibinfo  {journal} {Phys. Rev. B}\ }\textbf {\bibinfo {volume} {93}},\
  \bibinfo {pages} {075135} (\bibinfo {year} {2016})}\BibitemShut {NoStop}%
\bibitem [{\citenamefont {Witten}(2016)}]{Witten:2016ab}%
  \BibitemOpen
  \bibfield  {author} {\bibinfo {author} {\bibfnamefont {E.}~\bibnamefont
  {Witten}},\ }\href {\doibase 10.1103/RevModPhys.88.035001} {\bibfield
  {journal} {\bibinfo  {journal} {Rev. Mod. Phys.}\ }\textbf {\bibinfo {volume}
  {88}},\ \bibinfo {pages} {035001} (\bibinfo {year} {2016})}\BibitemShut
  {NoStop}%
\bibitem [{\citenamefont {Greiter}\ \emph {et~al.}(2007)\citenamefont
  {Greiter}, \citenamefont {Rachel},\ and\ \citenamefont
  {Schuricht}}]{Greiter:2007ac}%
  \BibitemOpen
  \bibfield  {author} {\bibinfo {author} {\bibfnamefont {M.}~\bibnamefont
  {Greiter}}, \bibinfo {author} {\bibfnamefont {S.}~\bibnamefont {Rachel}}, \
  and\ \bibinfo {author} {\bibfnamefont {D.}~\bibnamefont {Schuricht}},\
  }\href@noop {} {\bibfield  {journal} {\bibinfo  {journal} {Phys. Rev. B}\
  }\textbf {\bibinfo {volume} {75}},\ \bibinfo {pages} {060401} (\bibinfo
  {year} {2007})}\BibitemShut {NoStop}%
\bibitem [{\citenamefont {Greiter}\ and\ \citenamefont
  {Rachel}(2007)}]{Greiter:2007aa}%
  \BibitemOpen
  \bibfield  {author} {\bibinfo {author} {\bibfnamefont {M.}~\bibnamefont
  {Greiter}}\ and\ \bibinfo {author} {\bibfnamefont {S.}~\bibnamefont
  {Rachel}},\ }\href {https://link.aps.org/doi/10.1103/PhysRevB.75.184441}
  {\bibfield  {journal} {\bibinfo  {journal} {Phys. Rev. B}\ }\textbf {\bibinfo
  {volume} {75}},\ \bibinfo {pages} {184441} (\bibinfo {year}
  {2007})}\BibitemShut {NoStop}%
\bibitem [{\citenamefont {Takhtajan}(1982)}]{Takhtajan:1982aa}%
  \BibitemOpen
  \bibfield  {author} {\bibinfo {author} {\bibfnamefont {L.}~\bibnamefont
  {Takhtajan}},\ }\href@noop {} {\bibfield  {journal} {\bibinfo  {journal}
  {Phys. Lett. A}\ }\textbf {\bibinfo {volume} {87}},\ \bibinfo {pages} {479}
  (\bibinfo {year} {1982})}\BibitemShut {NoStop}%
\bibitem [{\citenamefont {Babujian}(1982)}]{Babujian:1982aa}%
  \BibitemOpen
  \bibfield  {author} {\bibinfo {author} {\bibfnamefont {H.}~\bibnamefont
  {Babujian}},\ }\href@noop {} {\bibfield  {journal} {\bibinfo  {journal}
  {Phys. Lett. A}\ }\textbf {\bibinfo {volume} {90}},\ \bibinfo {pages} {479}
  (\bibinfo {year} {1982})}\BibitemShut {NoStop}%
\bibitem [{\citenamefont {Andrei}\ and\ \citenamefont
  {Johannesson}(1984)}]{Andrei:1984aa}%
  \BibitemOpen
  \bibfield  {author} {\bibinfo {author} {\bibfnamefont {N.}~\bibnamefont
  {Andrei}}\ and\ \bibinfo {author} {\bibfnamefont {H.}~\bibnamefont
  {Johannesson}},\ }\href@noop {} {\bibfield  {journal} {\bibinfo  {journal}
  {Phys. Lett. A}\ }\textbf {\bibinfo {volume} {104}},\ \bibinfo {pages} {370}
  (\bibinfo {year} {1984})}\BibitemShut {NoStop}%
\bibitem [{\citenamefont {Johannesson}(1986)}]{Johannesson:1986aa}%
  \BibitemOpen
  \bibfield  {author} {\bibinfo {author} {\bibfnamefont {H.}~\bibnamefont
  {Johannesson}},\ }\href@noop {} {\bibfield  {journal} {\bibinfo  {journal}
  {Nucl. Phys. B}\ }\textbf {\bibinfo {volume} {270}},\ \bibinfo {pages} {235}
  (\bibinfo {year} {1986})}\BibitemShut {NoStop}%
\bibitem [{\citenamefont {Rachel}\ \emph {et~al.}(2009)\citenamefont {Rachel},
  \citenamefont {Thomale}, \citenamefont {F{\"u}hringer}, \citenamefont
  {Schmitteckert},\ and\ \citenamefont {Greiter}}]{Rachel:2009aa}%
  \BibitemOpen
  \bibfield  {author} {\bibinfo {author} {\bibfnamefont {S.}~\bibnamefont
  {Rachel}}, \bibinfo {author} {\bibfnamefont {R.}~\bibnamefont {Thomale}},
  \bibinfo {author} {\bibfnamefont {M.}~\bibnamefont {F{\"u}hringer}}, \bibinfo
  {author} {\bibfnamefont {P.}~\bibnamefont {Schmitteckert}}, \ and\ \bibinfo
  {author} {\bibfnamefont {M.}~\bibnamefont {Greiter}},\ }\href
  {https://link.aps.org/doi/10.1103/PhysRevB.80.180420} {\bibfield  {journal}
  {\bibinfo  {journal} {Phys. Rev. B}\ }\textbf {\bibinfo {volume} {80}},\
  \bibinfo {pages} {180420} (\bibinfo {year} {2009})}\BibitemShut {NoStop}%
\bibitem [{\citenamefont {Lajk{\'o}}\ \emph {et~al.}(2017)\citenamefont
  {Lajk{\'o}}, \citenamefont {Wamer}, \citenamefont {Mila},\ and\ \citenamefont
  {Affleck}}]{Lajko:2017aa}%
  \BibitemOpen
  \bibfield  {author} {\bibinfo {author} {\bibfnamefont {M.}~\bibnamefont
  {Lajk{\'o}}}, \bibinfo {author} {\bibfnamefont {K.}~\bibnamefont {Wamer}},
  \bibinfo {author} {\bibfnamefont {F.}~\bibnamefont {Mila}}, \ and\ \bibinfo
  {author} {\bibfnamefont {I.}~\bibnamefont {Affleck}},\ }\href@noop {}
  {\bibfield  {journal} {\bibinfo  {journal} {Nucl. Phys. B}\ }\textbf
  {\bibinfo {volume} {924}},\ \bibinfo {pages} {508} (\bibinfo {year}
  {2017})}\BibitemShut {NoStop}%
\bibitem [{\citenamefont {Dufour}\ \emph {et~al.}(2015)\citenamefont {Dufour},
  \citenamefont {Nataf},\ and\ \citenamefont {Mila}}]{Dufour:2015aa}%
  \BibitemOpen
  \bibfield  {author} {\bibinfo {author} {\bibfnamefont {J.}~\bibnamefont
  {Dufour}}, \bibinfo {author} {\bibfnamefont {P.}~\bibnamefont {Nataf}}, \
  and\ \bibinfo {author} {\bibfnamefont {F.}~\bibnamefont {Mila}},\ }\href
  {https://link.aps.org/doi/10.1103/PhysRevB.91.174427} {\bibfield  {journal}
  {\bibinfo  {journal} {Phys. Rev. B}\ }\textbf {\bibinfo {volume} {91}},\
  \bibinfo {pages} {174427} (\bibinfo {year} {2015})}\BibitemShut {NoStop}%
\bibitem [{\citenamefont {Lecheminant}(2015)}]{Lecheminant:2015aa}%
  \BibitemOpen
  \bibfield  {author} {\bibinfo {author} {\bibfnamefont {P.}~\bibnamefont
  {Lecheminant}},\ }\href@noop {} {\bibfield  {journal} {\bibinfo  {journal}
  {Nucl. Phys. B}\ }\textbf {\bibinfo {volume} {901}},\ \bibinfo {pages} {510}
  (\bibinfo {year} {2015})}\BibitemShut {NoStop}%
\bibitem [{\citenamefont {Chen}\ \emph {et~al.}(2013)\citenamefont {Chen},
  \citenamefont {Gu}, \citenamefont {Liu},\ and\ \citenamefont
  {Wen}}]{Chen:2013aa}%
  \BibitemOpen
  \bibfield  {author} {\bibinfo {author} {\bibfnamefont {X.}~\bibnamefont
  {Chen}}, \bibinfo {author} {\bibfnamefont {Z.-C.}\ \bibnamefont {Gu}},
  \bibinfo {author} {\bibfnamefont {Z.-X.}\ \bibnamefont {Liu}}, \ and\
  \bibinfo {author} {\bibfnamefont {X.-G.}\ \bibnamefont {Wen}},\ }\href
  {\doibase 10.1103/PhysRevB.87.155114} {\bibfield  {journal} {\bibinfo
  {journal} {Phys. Rev. B}\ }\textbf {\bibinfo {volume} {87}},\ \bibinfo
  {pages} {155114} (\bibinfo {year} {2013})}\BibitemShut {NoStop}%
\bibitem [{App()}]{Append}%
  \BibitemOpen
  \href@noop {} {\emph {\bibinfo {title} {{See Supplemental
  Materials.}}}}\BibitemShut {Stop}%
\bibitem [{\citenamefont {Affleck}(1988)}]{Affleck:1988aa}%
  \BibitemOpen
  \bibfield  {author} {\bibinfo {author} {\bibfnamefont {I.}~\bibnamefont
  {Affleck}},\ }\href@noop {} {\bibfield  {journal} {\bibinfo  {journal} {Nucl.
  Phys. B}\ }\textbf {\bibinfo {volume} {305}},\ \bibinfo {pages} {582}
  (\bibinfo {year} {1988})}\BibitemShut {NoStop}%
\bibitem [{Note3()}]{Note3}%
  \BibitemOpen
  \bibinfo {note} {There is no 't Hooft anomaly for $PSU(N)$
  itself}\BibitemShut {NoStop}%
\bibitem [{\citenamefont {Naculich}\ and\ \citenamefont
  {Schnitzer}(1990)}]{Naculich-Schnitzer90}%
  \BibitemOpen
  \bibfield  {author} {\bibinfo {author} {\bibfnamefont {S.~G.}\ \bibnamefont
  {Naculich}}\ and\ \bibinfo {author} {\bibfnamefont {H.~J.}\ \bibnamefont
  {Schnitzer}},\ }\href@noop {} {\bibfield  {journal} {\bibinfo  {journal}
  {Nucl. Phys. B}\ }\textbf {\bibinfo {volume} {347}},\ \bibinfo {pages} {687}
  (\bibinfo {year} {1990})}\BibitemShut {NoStop}%
\bibitem [{\citenamefont {Kiritsis}\ and\ \citenamefont
  {Niarchos}(2011)}]{Kiritsis-Niarchos10}%
  \BibitemOpen
  \bibfield  {author} {\bibinfo {author} {\bibfnamefont {E.}~\bibnamefont
  {Kiritsis}}\ and\ \bibinfo {author} {\bibfnamefont {V.}~\bibnamefont
  {Niarchos}},\ }\href@noop {} {\bibfield  {journal} {\bibinfo  {journal}
  {JHEP}\ }\textbf {\bibinfo {volume} {04}},\ \bibinfo {pages} {113} (\bibinfo
  {year} {2011})}\BibitemShut {NoStop}%
\bibitem [{\citenamefont {Schulz}(1986)}]{Schulz:1986aa}%
  \BibitemOpen
  \bibfield  {author} {\bibinfo {author} {\bibfnamefont {H.~J.}\ \bibnamefont
  {Schulz}},\ }\href {https://link.aps.org/doi/10.1103/PhysRevB.34.6372}
  {\bibfield  {journal} {\bibinfo  {journal} {Phys. Rev. B}\ }\textbf {\bibinfo
  {volume} {34}},\ \bibinfo {pages} {6372} (\bibinfo {year}
  {1986})}\BibitemShut {NoStop}%
\bibitem [{\citenamefont {Ziman}\ and\ \citenamefont
  {Schulz}(1987)}]{Ziman:1987aa}%
  \BibitemOpen
  \bibfield  {author} {\bibinfo {author} {\bibfnamefont {T.}~\bibnamefont
  {Ziman}}\ and\ \bibinfo {author} {\bibfnamefont {H.~J.}\ \bibnamefont
  {Schulz}},\ }\href {https://link.aps.org/doi/10.1103/PhysRevLett.59.140}
  {\bibfield  {journal} {\bibinfo  {journal} {Phys. Rev. Lett.}\ }\textbf
  {\bibinfo {volume} {59}},\ \bibinfo {pages} {140} (\bibinfo {year}
  {1987})}\BibitemShut {NoStop}%
\bibitem [{\citenamefont {{Duivenvoorden}}\ and\ \citenamefont
  {{Quella}}(2013)}]{Duivenvoorden-Quella12}%
  \BibitemOpen
  \bibfield  {author} {\bibinfo {author} {\bibfnamefont {K.}~\bibnamefont
  {{Duivenvoorden}}}\ and\ \bibinfo {author} {\bibfnamefont {T.}~\bibnamefont
  {{Quella}}},\ }\href {\doibase 10.1103/PhysRevB.87.125145} {\bibfield
  {journal} {\bibinfo  {journal} {\prb}\ }\textbf {\bibinfo {volume} {87}},\
  \bibinfo {eid} {125145} (\bibinfo {year} {2013})},\ \Eprint
  {http://arxiv.org/abs/1206.2462} {arXiv:1206.2462 [cond-mat.str-el]}
  \BibitemShut {NoStop}%
\bibitem [{\citenamefont {Sengupta}(1997)}]{Sengupta:1997aa}%
  \BibitemOpen
  \bibfield  {author} {\bibinfo {author} {\bibfnamefont {A.}~\bibnamefont
  {Sengupta}},\ }\href@noop {} {\emph {\bibinfo {title} {Gauge theory on
  compact surfaces}}},\ Vol.\ \bibinfo {volume} {600 0821804847}\ (\bibinfo
  {publisher} {American Mathematical Soc.},\ \bibinfo {year}
  {1997})\BibitemShut {NoStop}%
\bibitem [{\citenamefont {Steenrod}(1944)}]{Steenrod:1944aa}%
  \BibitemOpen
  \bibfield  {author} {\bibinfo {author} {\bibfnamefont {N.}~\bibnamefont
  {Steenrod}},\ }\href@noop {} {\bibfield  {journal} {\bibinfo  {journal} {Ann.
  Math.}\ ,\ \bibinfo {pages} {294}} (\bibinfo {year} {1944})}\BibitemShut
  {NoStop}%
\bibitem [{\citenamefont {Witten}(2018)}]{Witten:2018aa}%
  \BibitemOpen
  \bibfield  {author} {\bibinfo {author} {\bibfnamefont {E.}~\bibnamefont
  {Witten}},\ }\href {\doibase https://doi.org/10.1016/j.aim.2017.06.021}
  {\bibfield  {journal} {\bibinfo  {journal} {Adv. Math.}\ }\textbf {\bibinfo
  {volume} {327}},\ \bibinfo {pages} {624} (\bibinfo {year}
  {2018})}\BibitemShut {NoStop}%
\end{thebibliography}%

\clearpage

\appendix
\begin{widetext}
\section{SUPPLEMENTAL MATERIALS}

\section{Mixed $PSU(N)-\mathbb{Z}$ anomaly}
\label{Mixed anomaly}

The correspondence between mixed $PSU(N)-\mathbb{Z}$ anomalies and elements of $H^2(PSU(N), U(1))$ can be seen in a more explicit way, based on a similar argument in
\cite{Metlitski-Thorngren17} for the case of $PSU(2) = SO(3)$ symmetry,
when we write down the bulk effective action of a (2+1)d SPT phase (protected by both $PSU(N)$ and $\mathbb{Z}$ symmetries) coupled to a $PSU(N)\times\mathbb{Z}$ background field:
\begin{align}
\label{3d_bulk_action}
S_{\mathrm{bulk}}=\frac{2\pi i }{N}\int_{X_3}af^*(\alpha).
\end{align}
Here $X_3$ is the bulk three-manifold, $a\in H^1(X_3, \mathbb{Z})\cong\mathbb{Z}$ is the background $\mathbb{Z}$ gauge field, $f$ is a classifying map from $X_3$ to $BPSU(N)$ -- the classifying space of $PSU(N)$, and 
$\alpha\in H^2(BPSU(N), U(1))= H^2(PSU(N), U(1))$ 
(so the pullback class $f^*(\alpha)\in H^2(X_3, \mathbb{Z}_N)\cong\mathbb{Z}_N$ is a characteristic class -- a generalization of the second Stiefel-Whitney class in the case $N=2$ -- of the background $PSU(N)$ bundle over $X_3$).
Then, based on the expression of $S_{\mathrm{bulk}}$ as well as on the fact that elements of $H^2(PSU(N), U(1))$ are represented by projective classes of $SU(N)$ representations, 
which correspond to numbers of YT boxes in $SU(N)$ representations 
\cite{Duivenvoorden-Quella12},
we conclude that any mixed anomaly of $PSU(N)\times\mathbb{Z}$ can be represented (by the bulk-boundary correspondence) by a $\mathbb{Z}_N$ number
\begin{align}
\label{mixed_anomaly_A1}
b_{\rho} \mod N
\end{align}
associated to some (linear) representation $\rho$ of $SU(N)$, where $b_{\rho}$ is the number of YT boxes in $\rho$.

\section{Constraint on the ground-state degeneracy by the LSM index}
Let us define the generator of $\mathbb{Z}^\text{trans}$ as $\mathscr{T}$. Correspondingly, $\mathscr{T}^p$ is the generator of $p\mathbb{Z}^\text{trans}$. 
The index related to $p\mathbb{Z}^\text{trans}$ is equal to $p\mathcal{I}_{N}$, which is trivial (that is, $p\mathcal{I}_N=0\mod N$) if and only if $p\in p_0\mathbb{N}$, where $p_0\equiv N/\text{gcd}(\mathcal{I}_{N},N)$. 
It means that the gapped phase is permitted to have no SSB in $p\mathbb{Z}^\text{trans}$ only if $p\in p_0\mathbb{N}$, while $p\mathbb{Z}^\text{trans}$ must be spontaneously broken if $p\notin p_0\mathbb{N}$ when gapped by a symmetry-respecting perturbation. 

As the first case that $\mathcal{I}_{N}= 0\mod N$, we can see that $p\mathbb{Z}^\text{trans}$ is not necessarily spontaneously broken for any integer $p\in\mathbb{N}$. Therefore, a trivial symmetric gapped phase is consistent with the imposed symmetry and the ground-state degeneracy $N_\text{GSD}\in\mathbb{N}$, which is not restricted. 

On the other hand, if $\mathcal{I}_{N}\neq 0\mod N$, given by a fixed ground state $|\text{G.S.}\rangle_0$, $\mathscr{T}^m|\text{G.S.}\rangle_0$ is also a ground state but not necessarily differs from $|\text{G.S.}\rangle_0$ since the spin-chain Hamiltonian is translationally invariant, We claim that $\{\mathscr{T}^m|\text{G.S.}\rangle_0:m=1,2,\cdots,p_0\}$ consists of exactly $p_0$ of distinct quantum states. If not so, let us assume, without loss of generality, that $1\leq m_1<m_2\leq p_0$ such that $\mathscr{T}^{m_2}|\text{G.S.}\rangle_0=\mathscr{T}^{m_1}|\text{G.S.}\rangle_0$ up to a trivial phase factor. Then acting on both side by $\mathscr{T}^{-m_1}$, we obtain that $\mathscr{T}^{m_2-m_1}|\text{G.S.}\rangle_0=|\text{G.S.}\rangle_0$, which means no SSB for $(m_2-m_1)\mathbb{Z}^\text{trans}$. This implies $(m_2-m_1)\in p_0\mathbb{N}$, which contradicts with the fact that $0<(m_2-m_1)<p_0$. 

Consequently, $p_0$ is the lower bound of ground-state degeneracy of gapped phases and the translation operator $\mathscr{T}$ must be acted $p_0\mathbb{N}$ times on any ground state to transform back to itself again. Assume ${p_0w_1}$ and $p_0w_2$ are, respectively, the minimal times to translate $\mathscr{T}^{m_1}|\text{G.S.}\rangle_0$ and $\mathscr{T}^{m_2}|\text{G.S.}\rangle_0$ back to themselves. Specifically, $w_{m_2}$ is the minimal positive integer such that $\mathscr{T}^{p_0w_{m_2}}\mathscr{T}^{m_{2}}|\text{G.S.}\rangle_0=\mathscr{T}^{m_{2}}|\text{G.S.}\rangle_0$. Then we act on both sides of $\mathscr{T}^{p_0w_{m_1}}\mathscr{T}^{m_{1}}|\text{G.S.}\rangle_0=\mathscr{T}^{m_{1}}|\text{G.S.}\rangle_0$ by $\mathscr{T}^{m_2-m_1}$ which implies $\mathscr{T}^{p_0w_{m_1}}\mathscr{T}^{m_{2}}|\text{G.S.}\rangle_0=\mathscr{T}^{m_{2}}|\text{G.S.}\rangle_0$. Thus, by definition of $w_{m_2}$, we obtain $w_{m_1}\geq w_{m_2}$. Similarly, $w_{m_1}\leq w_{m_2}$, which means $w_{m_1}=w_{m_2}$ is equal to a non-universal integer $w$ determined by detailed information of the system. It can then be concluded that the ground-state degeneracy is restricted to be a multiple of $p_0$, namely 
\begin{align}
\text{GSD}\in \frac{N}{\text{gcd}(\mathcal{I}_{N},N)}\mathbb{N}, 
\end{align}
and the ground state spontaneously breaks the translation symmetry down to a nonzero multiple of ${N}/{\text{gcd}(\mathcal{I}_{N},N)}$.

\section{Mixed $PSU(N)-\mathbb{Z}_n$ anomaly in $SU(N)_k$ WZW theory}
\label{Dirac index in QCD$_2$}
In this part, we characterize the mixed $PSU(N)-\mathbb{Z}_n$ anomaly in $SU(N)_k$ WZW theory on any orientable two-dimensional manifold $M_2$ with $\mathbb{Z}_n$ in the form of Eq.~(\ref{axial}). 
\subsection{Constrained Dirac fermion (CDF) theory}
It is useful to consider the following algebraic equivalence
\begin{eqnarray}
SU(N)_k\sim \frac{U(kN)_1}{U(k)}
\end{eqnarray}
which implies that, the left-hand side represented by $SU(N)_k$ WZW theory can be realized by $U(kN)_1$ CFT with $su(k)_N\oplus u(1)$ modded out. Since $u(kN)_1$ can be represented by the $k$-color $N$-flavor free complex fermion theory. Thus $SU(N)_k$ WZW theory is equivalent to the complex fermion theory coupled to color-$SU(k)$ and charge-$U(1)$ gauge fields, which is called $U(kN)/U(k)$ constrained Dirac fermion (CDF) theory. 

Furthermore, we need to couple to $U(kN)/U(k)$ CDF theory a background $PSU(N)$ gauge field to calculate the mixed $PSU(N)-\mathbb{Z}_n$ anomaly where the $Z_n$ transformation, on the fermionic side, takes the form of
\begin{eqnarray}
\label{translation}
\mathscr{T}:\,\,\psi_{c,\alpha}\rightarrow\exp\left(-i{\pi}\sigma_3m/{N}\right)\psi_{c,\alpha}
\end{eqnarray}
where the $z$-component Pauli matrix $\sigma_3$ carries the chirality index ``$\pm$'' of $\psi_{c,\alpha}\equiv[\psi_{+,c,\alpha},\psi_{-,c,\alpha}]$ and the $PSU(N)$ transformation acts on the $\alpha$-flavor degrees of freedom. Thus we obtain the gauged CDF (GCDF) theory: 
\begin{eqnarray}
\label{qcd}
Z_\text{CDF}[\mathscr{A}_{PSU(N)}]=\int\mathfrak{D}\mathscr{A}_{U(k)}\mathfrak{D}\left(\bar{\psi},\psi\right)\exp\left[i\int\bar{\psi}i\gamma^\mu({\partial_\mu}-{\mathscr{A}_\mu})\psi\right], \nonumber
\end{eqnarray}
by introducing the background gauge field $\mathscr{A}\in u(1)\oplus su(N)\oplus psu(k)=u(1)\oplus su(N)\oplus su(k)$. 

It should be noted that, although the gauge field $\mathscr{A}$ is the connection on principal bundle of the matrix group $PSU(N)\times U(k)$ where ``$\times$'' denotes a direct product.
We define the following groups $(G,*)$ and $(H,*)$ by direct product ``$\times$'': 
\begin{eqnarray}
G&\equiv&U(N)\times U(k);\\
H&\equiv&\left\{(u\mathbb{I}_{N\times N},\mathbb{I}_{k\times k})|u\in U(1)\right\}, 
\end{eqnarray}
where the group operation ``$*$'' is naturally defined as matrix product component by component. 

Since $PSU(N)=PU(N)=U(N)/U(1)$ where $U(1)$ represents a $U(1)$ center of $U(N)$ that consists of diagonal matrices, therefore, 
\begin{eqnarray}
PSU(N)\times U(k)&=&\frac{U(N)}{U(1)}\times U(k)
=\frac{U(N)\times U(k)}{H}
=G/H, 
\end{eqnarray}
where the quotient over $H$ is naturally induced from the inclusion $U(1)\rightarrow H$ that $U(1)$ elements are put into the first component of those of $H$.

\subsection{Fundamental group of $G/H$}
$H$ is a closed subgroup of Lie group $G$ which implies that $G$ is homeomorphic with a $H$-principal bundle over base space $G/H$.  Therefore, together with the pairwise-connectedness of $G$ and $G/H$, we have the following long exact sequence: 
\begin{eqnarray}
\ldots\rightarrow\pi_n(H)\xrightarrow{i_*}\pi_n(G)\xrightarrow{f_*}\pi_n(G/H)\xrightarrow{\partial_*}\pi_{n-1}(H)\rightarrow\ldots, \nonumber
\end{eqnarray}
where $\text{Im}(f_*)=\text{Ker}(\partial_*)$ and $\text{Ker}(f_*)=\text{Im}(i_*)$. $f_*$ is induced by the quotient or explicit direct product: 
\begin{eqnarray}
\label{quotient}
f(g_{U(N)},g_{U(k)})=g_{U(N)}\otimes g_{U(k)}, 
\end{eqnarray}
$i_*$ is induced by inclusion, and $\partial_*$ is induced by boundary mapping. Here ``$\oplus$'' denotes a tensor product. 

We take $n\equiv2l+1$ as general odd number which is $n=1$ or $l=0$ for our case: 
\begin{eqnarray}
\label{long}
\ldots\rightarrow\pi_{2l+1}(H)\xrightarrow{i_*}\pi_{2l+1}(G)\xrightarrow{f_*}\pi_{2l+1}(G/H)\xrightarrow{\partial_*}\pi_{2l}(H)\rightarrow\ldots, 
\end{eqnarray}
However, since $H$ is path-connected and $H\cong S^1$, we get
\begin{eqnarray}
\pi_{2l}(H)=0, 
\end{eqnarray}
which implies $\text{Ker}(\partial_*)=\pi_{2l+1}(G/H)$ and surjective $f_*$. Moreover, 
\begin{eqnarray}
\pi_{2l+1}(G/H)&=&\pi_{2l+1}(G)/\text{Ker}(f_*)
=\pi_{2l+1}(G)/\text{Im}(i_*), 
\end{eqnarray}
where ``$=$''s above represent group isomorphism. 

In addition, any transition function $\gamma(\theta)$ defining principal bundle $P(S^{2l+2},G/H)$ is classified by $\pi_{2l+1}(G/H)$ where the multi-component $\theta$ parametrizes the equator $S^{2l+1}$. However, due to the surjective $f_*$, there exists a ``loop'' or a transition function $\tilde{\gamma}(\theta)\in G$ defining $\tilde{P}(S^{2l+2},G)$ so that $f(\tilde{\gamma}(\theta))=\gamma(\theta)$. Therefore, for any $P(S^{2l+2},G/H)$ bundle, its transition function can be canonically decomposed into the elements in $G$: 
\begin{eqnarray}
t(\theta)=f(\tilde{\gamma}(\theta))=u(\theta)\otimes v(\theta)
\end{eqnarray}
with single-valued $u(\theta)$ and $v(\theta)$ which, by themselves, respectively, can define a $P(S^{2l+2},U(N),u)$ and a $P(S^{2l+2},U(k),v)$ bundles. 

As a special situation, in our case of $l=0$, more simplification can be done. First,
\begin{eqnarray}
\pi_{1}(G/H)&=&\pi_{1}(G)/\text{Im}(i_*)
=\pi_{1}(G)/i_*(\pi_1(H)). 
\end{eqnarray}
Generally, we have the following homeomorphism: 
\begin{eqnarray}
\label{homeo}
U(N)&\cong&U(1)\times SU(N), \\
U(1)&\equiv&\text{diag}[1,1,\cdots,1,\text{det}(U(N))]. 
\end{eqnarray}
Thus, since $\pi_1(SU(N))=\pi_1(SU(k))=0$, 
\begin{eqnarray}
\label{fundamental}
\pi_{1}(G/H)&\cong&\pi_{1}(G)/i_*(\pi_1(H))
=\pi_{1}(U(1))\times\pi_{1}(SU(N))\times\pi_1(U(1))\times\pi_{1}(SU(k))/\sim
=\pi_{1}(U(1))\times\pi_1(U(1))/\sim\nonumber\\
&=&(\mathbb{Z}\oplus\mathbb{Z})/\sim
\end{eqnarray}
where we have used that the fact that $\pi_{1}(SU(N))$ is trivial, and $\mathbb{Z}\oplus\mathbb{Z}$ is represented by an ordered integer pair $\langle n_1,n_2\rangle$: 
\begin{eqnarray}
\langle n_1,n_2\rangle&\equiv&\text{diag}[1,1,\cdots,1,\exp(in_1\theta)]_{N\times N}\times\text{diag}[1,1,\cdots,1,\exp(-in_2\theta)]_{k\times k}, 
\end{eqnarray}
where $\times$ by convention denotes direct product and all the ``$=$'' is understood as equivalence modulo homotopy. The relation ``$\sim$'' is defined by identification of $i_*(\pi_1(H))$ with the trivial element of $\pi_1(G)$. Since $\pi_1(H)\cong\mathbb{Z}$ has a generator as
\begin{eqnarray}
h\equiv(\exp(i\theta)\mathbb{I}_{N\times N},\mathbb{I}_{k\times k})
\end{eqnarray}
and, for any element $h^W\in\pi_1(H)$ with $W\in\mathbb{Z}$, 
\begin{eqnarray}
i_*(h^W)&=&i_*\left((\exp(iW\theta)\mathbb{I}_{N\times N},\mathbb{I}_{k\times k})\right)
=\langle WN,0\rangle, 
\end{eqnarray}
where ``$=$''s are equivalence modulo homotopy. 
It implies that $\langle n_1,n_2\rangle\sim\langle n'_1,n'_2\rangle$ if and only if
\begin{eqnarray}
\exists W\in\mathbb{Z}\text{ s.t. }n'_1=n_1+WN,\,n'_2=n_2, 
\end{eqnarray}
which means that $[n_1,n_2]$ lives on a strip, or that the equivalence relation can be equally imposed by restriction $1\leq n_1\leq N, n_2\in\mathbb{Z}$. Then we obtain the corresponding element generating $\pi_1(G/H)$ as 
\begin{eqnarray}
\gamma[n_1\text{ mod } N,n_2]&=&\text{diag}[1,1,\cdots,1,\exp(in_1\theta)]_{N\times N}\otimes\text{diag}[1,1,\cdots,1,\exp(-in_2\theta)]_{k\times k}
\end{eqnarray}
where we use the fact that ``$\cong$'' in the first line of Eq.~(\ref{fundamental}) is the isomorphism by $f_*$.

The above arguments can be equivalently stated that we can further decompose $\tilde{\gamma}(\theta)$ into an element in $U(1)\times U(1)\times SU(N)\times SU(k)$. Since $\pi_1(SU(N))=\pi_1(SU(k))=0$, we can further homotopically drag $\tilde{\gamma}(\theta)$ canonically to $\tilde{\gamma}_\text{a}(\theta)\in U(1)\times U(1)\subset U(1)\times U(1)\times SU(N)\times SU(k)$ by continuous $F(\theta,s)$ with $s\in[0,1]$: 
\begin{eqnarray}
F(\theta,s)=\left\{\begin{array}{cc}\tilde{\gamma}(\theta), &s=0\\\tilde{\gamma}_\text{a}(\theta), &s=1\end{array}\right., 
\end{eqnarray}
which naturally induces the homotopy of $\gamma(\theta)$ due to continuity of $f$: 
\begin{eqnarray}
f\circ F(\theta,s)=\left\{\begin{array}{cc}{\gamma}(\theta), &s=0\\\gamma_\text{a}(\theta), &s=1\end{array}\right., 
\end{eqnarray}
where $\gamma_a(\theta)\in U(1)\subset G/H$.

\subsection{Generalized clutching construction}

Any two-dimensional orientable manifold $M_2$ can be obtained by a quotient map $p:\,D\rightarrow M$ with the domain as a two-dimensional closed disk $D$ with $\partial D=S^1$ having an internal point $x_0\in D^\mathrm{o}$. For instance, a sphere can be obtained by pasting the left and the right halves of $\partial D$ and a torus $T^2$ can be given by identifying opposite edges of four arcs of $\partial D$. In general $p|_{D^\mathrm{o}}$ is an inclusion while $p|_{\partial D}$ defines a pasting rule of points on $\partial D$~\cite{Sengupta:1997aa}. 

Then, since $p$ is surjective, we can define two patches of $M_2$ by $M=U_1\cup U_2$ with
\begin{eqnarray}
U_1&\equiv&p(D-\{x_0\});\,\,U_2\equiv p(D-\partial D). 
\end{eqnarray}
We take any continuous function $\bar{\gamma}:\,D-\{x_0\}\rightarrow G/H$ and there exists a standard construction~\cite{Steenrod:1944aa} of a principal $G/H$-bundle $P$ with the following trivializations $\phi_i:U_i\times G/H\rightarrow P$ satisfying, for any $(p(x),g)\in(U_1\cap U_2)\times G/H$ $(g\in G/H)$, 
\begin{eqnarray}
\phi_1(p(x),\bar{\gamma}(x)g)=\phi_2(p(x),g), 
\end{eqnarray}
which means that $\bar{\gamma}\circ p(x)$ serves as a transition function on the intersection $U_1\cap U_2$. Since $G/H$ is pathwise-connected, the equivalence class of principal bundle $\{[P(M_2,G/H)]\}$ has a one-to-one correspondence with $\{[\bar{\gamma}|_{\partial D}]\}=\pi_{1}(G/H)$ for any two-dimensional orientable closed manifold $M_2$~\cite{Sengupta:1997aa}. Moreover, we can take a submanifold $S=S^1$ in the interior $(D-\{x_0\})^\mathrm{o}$ such that $S^1$ winds $x_0$ once. Then the straight-line homotopy from $\partial D$ to $S$ naturally induced an isomorphism between $[\bar{\gamma}|_{\partial D}]$ and $[\bar{\gamma}|_S]$. Therefore, we define $\gamma\equiv\bar{\gamma}|_{S}$ and obtain that $[P(M_2,G/H)]$ has a one-to-one correspondence with $[\gamma]$, or equivalently, $[\gamma]$ classifies $[P(M_2,G/H)]$. 

Together with the previous results on $\pi_{1}(G/H)$, we come to the conclusion that any $P(M_2,G/H)$ defined by $\gamma(\theta)$ is equivalent to some another $P_a(M_2,G/H)$ defined by $\gamma_a(\theta)$ which is diagonal matrices and satisfies $[\gamma_a]=[\gamma]$. It is characterized by an integer ordered pair: $\{\langle n_1,n_2\rangle|1\leq n_1\leq N, n_2\in\mathbb{Z}\}$
\begin{eqnarray}
\gamma_a(\theta)&=&\text{diag}[1,1,\cdots,1,\exp(in_1\theta)]_{N\times N}
\otimes\text{diag}[1,1,\cdots,1,\exp(-in_2\theta)]_{k\times k}, 
\end{eqnarray}
which explicitly means all the non-diagonal elements of $\gamma(\theta)$ can be canonically gauged out by a global gauge transformation on the manifold. 
Then
\begin{eqnarray}
\text{ind}(i\slashed{\triangledown})&=&\int_{M_2}\text{ch}_1(P)
=\text{Tr}\int_{p(S)}i[\gamma\circ p^{-1}]^{-1}d[\gamma\circ p^{-1}]/2\pi
=\text{Tr}\int_{S}i\gamma(\theta)^{-1}\partial_\theta\gamma(\theta)/2\pi
=\text{Tr}\int_{S}i\gamma_a(\theta)^{-1}\partial_\theta\gamma_a(\theta)/2\pi\nonumber\\
&=&kn_1-Nn_2, 
\end{eqnarray}
where we have used the fact that $p$ when its range is restricted on $p(S)$ while domain restricted on $S$ has an inverse denoted by $p^{-1}$. 
Therefore, the anomaly is characterized by
\begin{eqnarray}
&&\exp\left[2i\alpha\cdot\text{ind}(i\slashed{\triangledown})\right]
=\exp\left[i\frac{2\pi m}{N}(kn_1-Nn_2)\right]
=\exp\left[i\frac{2\pi m}{N}(kn_1)\right], 
\end{eqnarray}
where $\alpha=m/N$. As a trivial check, the above results, such as the index and the anomaly factor, are independent of the representation chosen in the equivalence class defined by the relation $\sim$. 

From this result, we can see the $U(k)$ gauge field does not contribute to the anomaly factor since $\exp\left[2i\alpha\cdot\text{ind}(i\slashed{\triangledown})\right]$ is independent of $n_2$ -- as can be expected. More essentially, global non-abelian gauge structures are fully classified by abelian sectors on two-dimensional orientable manifolds $M_2$. 
Regarding this fact, we can further relate the value of the Dirac index to $k$ times (the integral of) the cohomology class $w\in\ H^2(M_2, \mathbb{Z}_N)$ of the background $PSU(N)$ bundle, namely
\begin{align}
\text{ind}(i\slashed{\triangledown}[\mathscr{A}_{PSU(N)\times U(k)}]) 
&= \int_{M_2} c_1(P_{PSU(N)\times U(k)}) 
 = k \int_{M_2} c_1(P_{U(N)}) \mod N
= k \int_{M_2} w(P_{PSU(N)}) \mod N,
\end{align}
where the last equality holds for any $U(N)$ bundle $P_{U(N)}$ lifted from the $PSU(N)$ bundle $P_{PSU(N)}$
\cite{Witten:2018aa, Gaiotto:2017aa}.
Therefore, the axial anomaly depends only on the underlying background $PSU(N)$ gauge field the theory is coupled to -- as we expect. By Fujikawa's method, we eventually have an anomalous phase change, which can be extracted out the functional integral $\int\mathscr{A}_{U(k)}$ in Eq.~(\ref{qcd}), of the total partition function under the discrete axial transformation as
\begin{align}
Z_\text{CDF}[\mathscr{A}_{PSU(N)}]\rightarrow
\exp\left(\frac{2\pi i km}{N}\int_{M_2}w(P_{PSU(N)})\right)Z_\text{CDF}[\mathscr{A}_{PSU(N)}] .
\end{align}

\section{Derivation of the condition (\ref{matching_emergent})}
\label{Derivation of the condition}

Consider a translationally symmetric $SU(N)$ spin model with an LSM index $\mathcal{I}_N$. We suppose this model has an emergent $SU(N')_{k'}$ WZW critical theory with $N'>N$ and the $\mathbb{Z}$ symmetry (translation symmetry at low energy) specified by $g\rightarrow e^{2\pi im'/N'}g$. Then, from the definition of the LSM index, such a critical theory has a mixed $PSU(N)-\mathbb{Z}$ anomaly characterized by $\mathcal{I}_N$, or equivalently, by a phase factor $\exp(2\pi i\mathcal{I}_N/N)\in \mathbb{Z}_N$. On the other hand, while this WZW theory respects a larger vector $PSU(N')$ symmetry than $PSU(N)$ (the latter is a subgroup of the former), one can also compute the mixed anomaly between the $PSU(N')$ and the $\mathbb{Z}$ symmetries, which, as discussed in the main text, is represented by another phase factor $\exp(2\pi ik'm'/N')\in \mathbb{Z}_{N'}$. 

Now, take $p:=N'/\text{gcd}(N', k'm')$ identical copies of the $SU(N')_{k'}$ WZW theory and denote such a theory as $\mathcal{W}$. Obviously, $\mathcal{W}$ is free of the mixed $PSU(N')-\mathbb{Z}$ anomaly, as the associated anomaly phase factor $\exp(2\pi ipk'm'/N')$ is equal to 1. (Note that $p$ is the minimal integer for any anomaly-free theory constructed in this way). Moreover, there is also no mixed 't Hoof t anomaly for the $PSU(N)\times\mathbb{Z}$ symmetry of $\mathcal{W}$, because any subgroup of an anomaly-free symmetry group in a theory does not have an 't Hooft anomaly. Thus, we must have  
$\exp(2\pi ip\mathcal{I}_N/N)= 1$, namely
\begin{align}
\mathcal{I}_N\cdot\frac{N'}{\text{gcd}(N',k'm')}=0 \mod N.
\end{align}

\end{widetext}

\end{document}